\documentclass[sigconf]{acmart}

\AtBeginDocument{%
}

\usepackage{xspace}
\usepackage[acronym]{glossaries}

\newacronym{llm}{LLM}{large language model}
\newacronym{ea}{EA}{evolutionary algorithm}
\newacronym{iec}{IEC}{interactive evolutionary computation}
\newacronym{hci}{HCI}{human-computer interaction}
\newacronym{va}{VA}{visual analytics}
\newacronym{ai}{AI}{artificial intelligence}
\newacronym{ml}{ML}{machine learning}
\newacronym{dl}{DL}{deep learning}
\newacronym{oxc}{OXC}{optical circuit switch}
\newacronym{nl}{NL}{natural-language}
\newacronym{sa}{SA}{situation awareness}
\newacronym{loa}{LOA}{levels of automation}

\newcommand{\toolname}{EvoMaestro\xspace}
\newcommand{\etal}{et~al.}
\usepackage{pifont}
\newcommand{\circled}[1]{\ding{\numexpr171+#1\relax}}
\newcommand{\eg}{e.g.,}
\newcommand{\ie}{i.e.,}

\newcommand{\DR}[1]{\hyperref[dr:#1]{DR#1}}

\newif\ifhighlightcrrevisions
\highlightcrrevisionsfalse 
\definecolor{CRRevisionBlue}{RGB}{60,105,170}
\newcommand{\crrev}[1]{%
  \ifhighlightcrrevisions
  \textcolor{CRRevisionBlue}{#1}%
  \else
  #1%
\fi}

\definecolor{CaptionColor}{RGB}{89, 91, 97}
\newcommand*\captionID[1]{%
  \begingroup
  \setlength{\fboxrule}{1pt}%
  \setlength{\fboxsep}{1.2pt}%
  \fcolorbox{CaptionColor}{white}{\textcolor{CaptionColor}{\textbf{#1}}}%
  \endgroup}

\definecolor{FeatureColor}{RGB}{100,88,138}
\newcommand{\feature}[1]{\textit{#1}}
\newcommand{\featurebf}[1]{\textbf{\textit{#1}}}
\newcommand{\fsnum}{8}
\newcommand{\usnum}{12}

\copyrightyear{2026}
\acmYear{2026}
\setcopyright{cc}
\setcctype{by}
\acmConference[UIST '26]{The 39th Annual ACM Symposium on User Interface Software and Technology}{November 02--05, 2026}{Detroit, MI, USA}
\acmBooktitle{The 39th Annual ACM Symposium on User Interface Software and Technology (UIST '26), November 02--05, 2026, Detroit, MI, USA}
\acmDOI{10.1145/3830398.3830626}
\acmISBN{979-8-4007-2856-3/2026/11}

\acmSubmissionID{4349}

\begin{document}

\title{\toolname: Toward Interpretable and Steerable LLM-Driven Program Evolution}

\author{Feng Liang}
\email{feng011@e.ntu.edu.sg}
\affiliation{%
  \institution{Nanyang Technological University}
  \city{Singapore}
  \country{Singapore}
}

\author{Sizhe Cheng}
\email{sizhe003@e.ntu.edu.sg}
\affiliation{%
  \institution{Nanyang Technological University}
  \city{Singapore}
  \country{Singapore}
}

\author{Yikai Li}
\email{liyi0084@e.ntu.edu.sg}
\affiliation{%
  \institution{Nanyang Technological University}
  \city{Singapore}
  \country{Singapore}
}

\author{Ruijie He}
\email{herj25@mails.tsinghua.edu.cn}
\affiliation{%
  \institution{Tsinghua University}
  \city{Beijing}
  \country{China}
}

\author{Xiaolin Wen}
\email{xiaolin004@e.ntu.edu.sg}
\affiliation{%
  \institution{Nanyang Technological University}
  \city{Singapore}
  \country{Singapore}
}

\author{Yong Wang}
\email{yong.wang@ntu.edu.sg}
\affiliation{%
  \institution{Nanyang Technological University}
  \city{Singapore}
  \country{Singapore}
}

\renewcommand{\shortauthors}{Liang et al.}


\begin{abstract}
  Recent program evolution systems use large language models (LLMs) to generate and iteratively improve populations of programs, producing striking advances in mathematics, algorithm design, and scientific computing.
  Yet these systems largely operate as fully automated black boxes.
  \crrev{As populations grow, domain experts must make sense of the scores, code changes, reasoning, and algorithmic ideas across many programs, while current interfaces provide limited support for understanding or redirecting the evolution.
    We characterize this need to understand and steer populations of evolving algorithmic ideas as \emph{semantic oversight}.
  A formative study with \fsnum{} domain experts yields six design requirements for this emerging human-computer interaction problem.}
  We then propose a steerable program evolution framework that lets expert judgments shape subsequent evolution.
  Built on this framework, \toolname is an interactive visual analytics system that organizes evolution information from population overview to source code.
  It helps experts locate and compare noteworthy programs, guide evolution with natural language, combine promising ideas, and prune unproductive directions.
  \crrev{A seven-day system demonstration illustrates how an expert applied these capabilities throughout a long-running evolution process.}
  A within-subjects study with \usnum{} participants shows that \toolname improves users' understanding of evolution processes, reduces cognitive workload, and supports expert steering.
  \crrev{These findings suggest that preserving expert agency in open-ended LLM-driven search requires support for both informed judgment and the ability to shape subsequent automation.}
\end{abstract}

\ccsdesc[500]{Human-centered computing~Interactive systems and tools}
\ccsdesc[300]{Human-centered computing~Visualization systems and tools}

\keywords{LLM-driven program evolution, visual analytics}

\begin{teaserfigure}
  \centering
  \includegraphics[width=\linewidth]{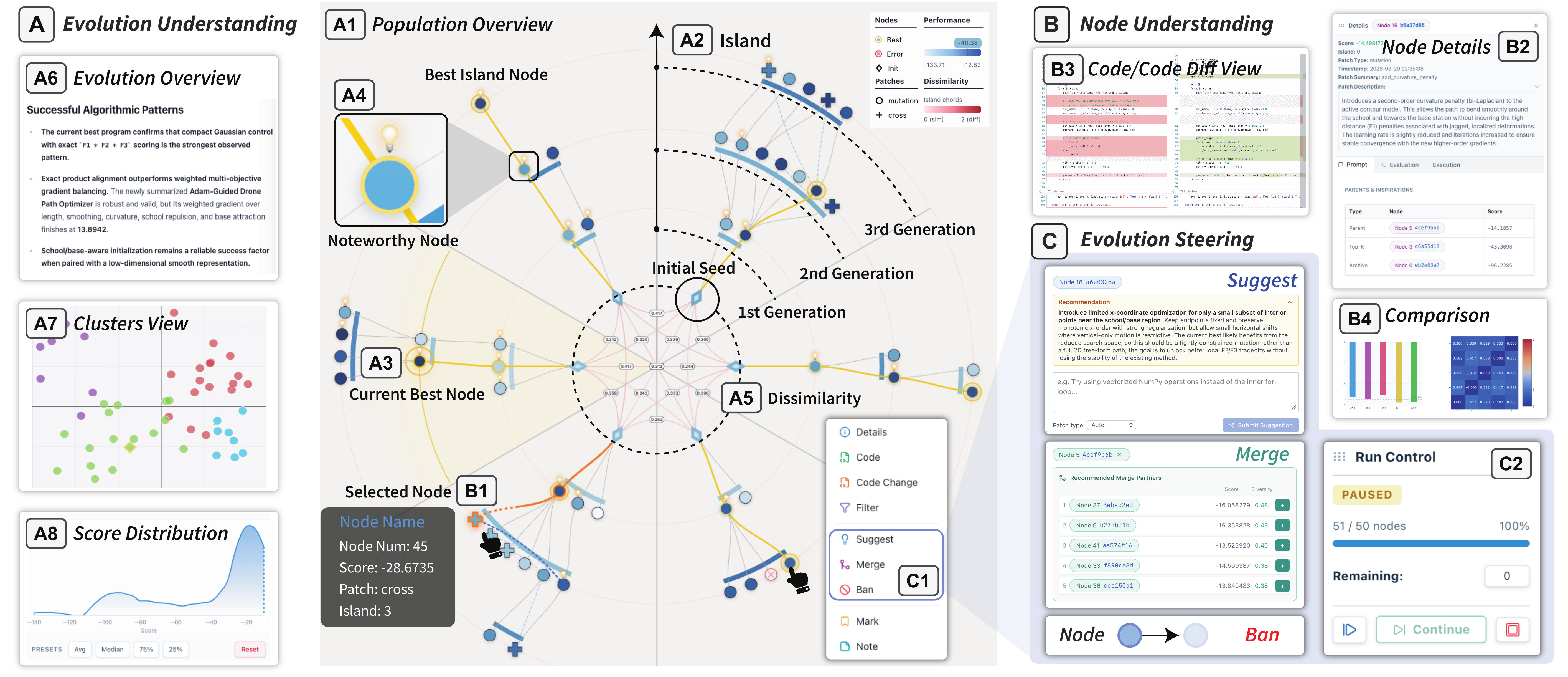}
  \caption{The \toolname interface supports LLM-driven program evolution through three stages: (A)~\textit{Evolution Understanding} overviews the population structure, performance, and strategies (A1-A8); (B)~\textit{Node Understanding} enables multi-level inspection of individual nodes, including metadata (B1, B2), code explanations (B3), and comparison (B4); and (C)~\textit{Evolution Steering} allows interactive interventions such as \textit{Suggest}, \textit{Merge}, \textit{Ban}, and \textit{pacing control} of the evolution process.
  }
  \Description{A wide annotated screenshot of the EvoMaestro interface, divided into three labeled sections.
    Section A, Evolution Understanding, occupies the left two-thirds and centers on a large radial tree visualization (A1, Population Overview) where nodes are connected by edges radiating outward from an Initial Seed at the center through concentric generation rings labeled 1st, 2nd, and 3rd Generation.
    Node fill color encodes performance; golden rings mark the current best node per island, lightbulb icons mark noteworthy nodes prioritized for expert review, and red circled X icons mark error nodes.
    Diamonds, circles, and plus signs encode initial, mutation, and cross patches, respectively.
    Dashed curved lines (A5, Dissimilarity chords) connect pairs of nodes.
    A legend in the upper right explains node and patch encodings, with a dissimilarity color scale ranging from dark blue (0, similar) to orange (2, different).
    Along the bottom of the tree, small inline sparkline charts (A2) show per-generation score trends on arc segments.
    To the left, panel A6 (Evolution Overview) displays an LLM-generated textual summary listing three successful algorithmic patterns as bullet points.
    Below it, panel A7 (Clusters View) shows a scatter plot of colored dots representing node clusters, and panel A8 (Score Distribution) shows a histogram of score values with preset filter buttons for Avg, Median, 75 percent, and 25 percent.
    Section B, Node Understanding, is in the upper right.
    Panel B1 shows a tooltip appearing on a selected node with its name, node number (45), score (-28.6735), patch type (cross), and island number (3).
    Panel B2 (Node Details) shows a detail panel listing the node's score, island, patch type, timestamp, patch summary, patch description, and a Parents and Inspirations table with parent, top-k, and archive entries showing node IDs and scores.
    It also has Prompt, Evaluation, and Execution tabs.
    Panel B3 (Code/Code Diff View) displays source code with syntax highlighting and line-level annotations from the Maestro Agent in a chat-like interface on the right side.
    Panel B4 (Comparison) shows a matrix heatmap of pairwise dissimilarity values among selected nodes, with cells colored from yellow-green (low) to dark purple (high).
    Section C, Evolution Steering, is in the lower right.
    Panel C1 shows a right-click context menu on a node with options: Details, Code, Code Change, Filter, Suggest, Merge, Ban, Mark, and Note.
    Below it, a Suggest modal displays a target node ID, a text area for the expert's recommendation with example guidance, a patch type selector, and a Submit Suggestion button.
    A Merge modal shows two selected node IDs, a list of recommended merge partners ranked by score and diversity, each with a blue add button.
    A Ban action is illustrated with an arrow from a node to a red ban icon.
    Panel C2 (Run Control) shows a paused evolution with a progress bar at 51 of 50 nodes (100 percent), a Remaining field set to 0, and Continue and stop buttons with playback controls.}
  \label{fig:interface}
\end{teaserfigure}

\maketitle



\section{Introduction}
\label{sec:introduction}


Advances in \acrfullpl{llm} have transformed how programmers write code, from inline completions to multi-file edits guided by \gls{nl} intent~\cite{vaithilingam_expectation_2022, barke_grounded_2023}.
Recent systems (\eg{}\cite{romera-paredes_mathematical_2024, novikov_alphaevolve_2025, lange_shinkaevolve_2025}) push this further: instead of assisting programmers, \glspl{llm} serve as the engine of evolutionary search over programs.
They maintain a \emph{population} of candidate programs and iteratively improve them over successive generations: an \gls{llm} reads select programs, reasons about their algorithmic strategies, and proposes modifications, while an evaluation function scores the results and feeds them back into the population.
This paradigm has produced striking results, including novel mathematical findings~\cite{romera-paredes_mathematical_2024, georgiev_mathematical_2025}, improvements on long-standing algorithms~\cite{novikov_alphaevolve_2025}, and new activation functions in \gls{dl}~\cite{vitvitskyi_mining_2026}.
As open-source frameworks~\cite{lange_shinkaevolve_2025, sharma_openevolve_2025} lower the entry barrier, \gls{llm}-driven program evolution is becoming a powerful tool for researchers across fields.


However, these systems operate as fully automated black boxes, lacking transparency and controllability during execution.
Once a run is launched, experts have limited ability to monitor, understand, or steer the evolution process.
This poses two key challenges (\textbf{C1}, \textbf{C2}).
First, experts struggle to \emph{understand} the evolution (\textbf{C1}): a run produces a rapidly growing tree of program variants, each carrying scores, code diffs, and error logs, and the sheer volume overwhelms human inspectors.
Unlike classical \glspl{ea}, \gls{llm}-driven evolution generates intentional \emph{code edits} accompanied by rich metadata (\eg{} reasoning chains and strategy descriptions), but existing interfaces (\eg{} ShinkaEvolve UI~\cite{lange_shinkaevolve_2025}) fail to effectively surface this information for analysis.
Second, experts lack mechanisms to \emph{steer} the evolution (\textbf{C2}): current systems do not support pausing, redirecting, combining solutions, or pruning unproductive lineages during execution.
Although experts possess domain knowledge that could redirect unproductive search directions and amplify promising ones, there is no effective channel to incorporate such human input into the evolution process.


To address these challenges, we present a novel interactive visual analytics system for interpretable and steerable \gls{llm}-driven program evolution.
Specifically, we first conducted a formative study with \fsnum{} domain experts to derive six concrete design requirements (\S~\ref{sec:formative-study}).
Informed by these requirements, we propose a steerable program evolution framework that extends the standard \gls{llm}-driven evolution process by enabling expert interventions (\eg{} manually guiding program generation) during evolution to support steering (\textbf{C2}) and exposing evolution details (\eg{} population structure, performance, and strategies) to support interpretation (\textbf{C1}).
Built on this framework, the \toolname interface incorporates a novel visual design of population overview and multiple coordinated analytical views to further support both intuitive evolution understanding (\textbf{C1}) and smooth expert steering interactions (\textbf{C2}).
The interface organizes evolution information through progressive disclosure from population-level overviews down to \gls{llm}-generated line-level code annotations, enabling experts to build situational awareness without being overwhelmed by raw code and scores.
\crrev{We demonstrate \toolname on a real-world optimization task} and evaluate it through an in-depth user study with 12 participants against the state-of-the-art baseline (\ie{} ShinkaEvolve Web UI~\cite{lange_shinkaevolve_2025}).
The results show that \toolname improves users' understanding of the evolution process, reduces cognitive workload, and supports effective expert steering.

In summary, we make the following contributions:

\begin{itemize}
  \item \crrev{We characterize \emph{semantic oversight} of \gls{llm}-driven program evolution as an emerging \gls{hci} problem: experts must understand and steer populations of evolving algorithmic ideas.} A formative study with \fsnum{} scientists and engineers yields six design requirements.
  \item We propose a steerable program evolution framework and \toolname\footnote{Opensourced on GitHub (\url{https://github.com/ifsheldon/EvoMaestro})}. The framework lets expert judgments shape subsequent evolution process; the interface helps experts locate noteworthy nodes and inspect strategies from the population level to source code.
  \item \crrev{We demonstrate sustained expert steering on a real-world optimization task and evaluate \toolname through a within-subjects study with \usnum{} participants.} The study shows that \toolname significantly improves experts' understanding of the evolving population, enables effective expert-guided steering, and reduces cognitive workload. \crrev{We distill two lessons for preserving expert agency: support informed judgment and action on that judgment.}
\end{itemize}


\section{Related Work}
\label{sec:related-work}
Our work relates to existing research on LLM-Driven Program Evolution, Visualization for Evolutionary Algorithms, and Human Oversight of Iterative Program Generation.

\subsection{LLM-Driven Program Evolution}

Recent studies use \glspl{llm} as mutation operators in evolution processes, generating code variants through prompting rather than syntactic manipulation.
FunSearch~\cite{romera-paredes_mathematical_2024} demonstrated that \gls{llm}-driven evolution could discover novel mathematical constructs.
AlphaEvolve~\cite{novikov_alphaevolve_2025} scaled up this approach for scientific and algorithmic discovery.
DeepEvolve~\cite{liu2025scientificalgorithmdiscoveryaugmenting} extends AlphaEvolve with deep research capabilities.
ShinkaEvolve~\cite{lange_shinkaevolve_2025} improves the search efficiency.
Open-source implementations~\cite{sharma_openevolve_2025, liu_llm4ad_2026, lange_shinkaevolve_2025} have broadened the access.
These systems are already producing real scientific impact~\cite{vitvitskyi_mining_2026, novikov_alphaevolve_2025, romera-paredes_mathematical_2024}.
\crrev{Although they generate code, their application domains are diverse: code can represent candidate solutions to problems in mathematics, computational chemistry, physics, biology and other fields.
  This approach applies when solutions can be (1) expressed as executable programs and (2) evaluated automatically through safe execution using a computable, sufficiently reliable metric.
Its potential users therefore include scientists and engineers who can judge domain-specific ideas but may need support to connect those ideas with generated code and evaluation results.}
Existing program evolution systems primarily operate as automated pipelines and provide limited support for expert steering during a run.
\toolname builds directly on these systems, providing an interactive layer that makes the evolution process legible and steerable by domain experts.





\subsection{Visualization for Evolutionary Algorithms}

For traditional \glspl{ea}, early work has visualized population dynamics~\cite{wu_visual_1999, kerren_eavis_2005}, fitness landscapes, multi-objective trade-offs~\cite{ma_paretolens_2025,he_visualization_2016,zhen_objective_2020,nagar_interpretable_2021, ibrahim_3dradvis_2016, daniels_properties_2012, yuan_dimension_2013} and neuroevolution trajectories~\cite{wang_vine_2018, boukhelifa_evolutionary_2013}.
NAS-Navigator~\cite{tyagi_nas-navigator_2023} and VisEvol~\cite{chatzimparmpas_visevol_2021} provide visual steering with domain knowledge injection and real-time intervention in neural architecture search, representing the closest design precedents to our work. These systems operate on search spaces where solutions are numerical vectors, modular components, or architectural blocks.
\Gls{llm}-driven program evolution differs from traditional \glspl{ea} in that each mutation is a \emph{reasoned code edit} accompanied by rich metadata (\eg{} reasoning chains, \gls{llm}-generated summaries) with no analog in classical \glspl{ea}, and code diffs are structurally complex. ShinkaEvolve~\cite{lange_shinkaevolve_2025} and OpenEvolve~\cite{sharma_openevolve_2025} provide a web UI for visualizing their evolutions, but they are limited to post-hoc dashboards and do not support real-time interaction or steering. LLM4AD~\cite{liu_llm4ad_2026} provides pre-run configurations and simple control (\eg{} start, resume) over an evolution on its UI. In our work, we not only visualize the evolution process but also enable real-time steering.

\subsection{Human Oversight of Iterative Program Generation}

Existing \gls{hci} research on human-AI collaboration in programming primarily studies a user working with one assistant or one active artifact through code suggestions and chat-based editing~\cite{vaithilingam_expectation_2022, barke_grounded_2023, mozannar_reading_2024, liang_large-scale_2024}.
Recent taxonomies of human-\gls{llm} roles~\cite{chappidi2026doeswhatarchetypesroles} also focus on dyadic archetypes.
\crrev{By contrast, in program evolution, the expert oversees persistent alternatives and shifts from co-author to curator and strategist: the expert must allocate attention, compare strategies, inspect evidence, follow lineages, and decide where to intervene.
  Mixed-initiative program synthesis provides a closer precedent.
  ASSUAGE~\cite{hu_assuage_2021} exposes parallel assembly-synthesis instances and lets users guide them through a predefined vocabulary of constraints and decompositions.
  Our system works with open-ended algorithmic ideas: \gls{llm}-generated programs can introduce strategies outside predefined categories, and experts can describe suggestions and new directions in \gls{nl}.
Steering such open-ended ideas requires \emph{semantic oversight}: experts must understand the strategies represented by generated programs, compare them across lineages, check them against code and evaluation results, decide which to preserve, combine, prune, or redirect, and monitor their effects.}

Several \gls{hci} frameworks help organize these needs.
Endsley's \gls{sa} framework~\cite{endsley_toward_1995} describes how operators perceive a system's state, comprehend what it means, and project what may happen next.
Sheridan's \gls{loa}~\cite{sheridan_telerobotics_1992} and Issak~\etal{}'s trajectory of control~\cite{controlisatrajectory} describe movement between passive monitoring and active intervention.
Pareek~\etal{}~\cite{pareek_sensemaking_2026} show that multi-agent interfaces must balance useful process visibility and trust against information overload.
\crrev{Our system supports semantic oversight through population-level attention routing, multi-level inspection of algorithmic strategies, and interventions that shape future generations.}


\section{Background: LLM-Driven Program Evolution}
\label{sec:background}

This section introduces the key concepts and terminology of \gls{llm}-driven program evolution.
All existing systems share a four-stage loop (Fig.~\ref{fig:framework}, Bottom Row), which continues until a termination condition is met, such as a time budget, a target score, or a maximum number of nodes.
In all existing systems, the loop is fully automatic.

\textbf{\circled{1} Population Database.}
An evolution maintains a \emph{population} of candidate programs stored in a population database.
Each candidate is represented as a \emph{node} in an evolution tree, carrying the program's source code together with metadata such as performance scores, code diffs, and \gls{llm} reasoning chains.
In many systems~\cite{romera-paredes_mathematical_2024, novikov_alphaevolve_2025,lange_shinkaevolve_2025}, the population is partitioned into \emph{islands}, sub-populations that evolve independently. Optionally, \emph{migrations} move nodes among islands.
These mechanisms, inspired by habitat isolation in nature, help maintain program diversity and improve performance.

\textbf{\circled{2} Parent Selection.}
In each generation, the system selects one or more nodes from the population as \emph{parents}, the starting points for producing the next generation of candidates.
The automatic and probabilistic selection typically favors nodes with higher scores, ensuring that good nodes are more likely to be refined further.

\textbf{\circled{3} Node Generation.}
New nodes are created by prompting an \gls{llm} to modify the selected parents.
In a \emph{mutation}, the \gls{llm} reads a single parent, reasons about its algorithm, and proposes a targeted code edit.
In a \emph{crossover}, the \gls{llm} reads two or more parents and combines their ideas into a new program.
In both cases, the \gls{llm} produces a \emph{patch} (code edits) applied to generate the new node, optionally with a \emph{patch description} summarizing the intent in \gls{nl}.
Unlike classical \glspl{ea}, where mutations are random perturbations, each \gls{llm}-generated patch is deliberate and semantically meaningful.

\textbf{\circled{4} Node Evaluation.}
An \emph{evaluation function} executes the newly generated program and assigns one or more numeric scores measuring solution quality.
This function is provided by the user and measures domain-specific metrics (\eg{} prediction accuracy and latency).
The evaluation stage can also generate additional metadata. For example, ShinkaEvolve~\cite{lange_shinkaevolve_2025} computes an embedding vector from each node's code for similarity comparison across the population.
The evaluated node, together with its scores and metadata, is then added back to the population database, and the loop repeats.

\section{Formative Study}
\label{sec:formative-study}
To inform the design of \toolname, we conducted a formative study with engineers and researchers who regularly solve optimization problems and could benefit from \gls{llm}-driven evolution.
This study aimed to understand three key aspects:
(1)~how they approach iterative algorithm improvement with \glspl{llm}, (2)~what information they seek when examining an \gls{llm}-driven evolution run, and (3)~what intervention capabilities they expect to effectively steer such processes. We used the web UI of ShinkaEvolve~\cite{lange_shinkaevolve_2025} as the baseline system for this study, as it is one of the most advanced open-source \gls{llm}-driven program evolution tools~\cite{sharma_openevolve_2025, liu_llm4ad_2026}.

\subsection{Participants and Procedures}
\label{sec:fs-method}

We recruited eight participants (\textbf{P1-P8}; 2 female, 6 male) from universities and companies across three countries via direct outreach and referrals (see Appendix~\ref{sec:appendix-fs-participants}).
All participants have substantial experience in algorithm design or optimization, including software engineers (P1, P2, P6) and graduate students (P3-P5, P7, P8) in \gls{dl} and mathematics.
All are proficient in reading code and regularly use \gls{llm}-based programming tools, with four (P1, P4-P6) having more than five years of competitive programming experience.
We did not require prior experience in \gls{ea}, as the target users are domain experts who seek to apply \gls{ea} techniques to their problems, rather than \gls{ea} specialists.
\crrev{Before the participant interviews, we collaborated with E1, a co-author and domain expert. He is a PhD student in theoretical computer science with six years of competitive programming experience and had extensive prior experience with the baseline system. We interviewed him about his experience with and feedback on the baseline. Based on his feedback and our literature review, the research team developed the two design probes. As E1 was a co-author, we did not include him among the recruited participants.}

We conducted a semi-structured interview with each participant, with each session lasting approximately 50 to 60 minutes (see Appendix~\ref{sec:appendix-fs-procedure} for details).
The two design probes covered alternative visualizations of evolution results and interaction mechanisms for steering an evolution process.
In each session, participants explored a pre-generated ShinkaEvolve~\cite{lange_shinkaevolve_2025} evolution with 50 nodes using a think-aloud protocol, followed by semi-structured questions. We then presented the probes and gathered feedback.


\subsection{Findings and Design Requirements}
\label{sec:fs-findings}

We conducted a thematic analysis following Braun and Clarke~\cite{braun_using_2006}.
\label{sec:fs-analysis}
The first author performed iterative open coding on interview notes, and themes were refined through team discussion.
Our analysis revealed six themes, which we distilled into six design requirements.
We organize them into two groups to reflect the two core problems identified from interview feedback:
\emph{information overload} (\DR{1}--\DR{4}) and \emph{lack of human-in-the-loop interaction} (\DR{5}--\DR{6}).

\subsubsection{Problem 1: Information Overload}

Participants reported being overwhelmed by the volume of information on the interface, yet paradoxically found it insufficient for understanding what they cared about most: what algorithmic strategies the nodes were using, which ones were promising, and how the search was progressing.

\textbf{DR1: Surface algorithmic intent of node generation.}\label{dr:1} Participants consistently reported that raw code and scores did not convey the algorithmic strategy behind each node.
P1 noted: \textit{``there is lots of information, but none of it helps me understand [what is going on].''}
While the baseline did include patch descriptions, they were buried and easy to miss. When participants found them, both P1 and P6 identified them as the most helpful information.
E1 exhibited a telling workaround: rather than reading code directly, he routinely copied it into ChatGPT to ask about the algorithmic idea.
The system should present the \emph{idea} behind each node, not just its code and score, in a way that is immediately visible and scannable.

\textbf{DR2: Support multi-level comparison between nodes.}\label{dr:2}
Comparing nodes was a major pain point.
P1 reported that \textit{``comparing code is tedious, and it's hard to compare the differences in thinking between solutions,''} and noted that \textit{``it's hard to tell at a glance whether performance improved or regressed''} when the parent and child had similar scores.
The baseline only supported comparing a node with its parent; E1 resorted to manually writing down scores for comparison.
Participants wanted comparison at multiple levels: across islands (P5), side-by-side code diffs (P1, P6), and strategy-level clustering (P5).
E1 likewise wanted comparison across islands.
The system should enable comparison at multiple abstraction levels.

\textbf{DR3: Provide population-level overviews.}\label{dr:3}
Multiple participants desired a global overview beyond the tree structure.
P1 and P6 articulated this clearly: \textit{``it would be best to have a global summary, like a report, summarizing what directions all these nodes explored.''}
P5 found it \textit{``difficult to find the top-K nodes.''} E1 asked for the global average score, which was not available.
The system should aggregate individual node information into overviews.

\textbf{DR4: Prioritize decision-relevant information.}\label{dr:4}
Even when relevant information was present, participants struggled to focus on what mattered.
P7 and P8 stated: \textit{``too much information with no focus.''}
The baseline mixed global statistics and local node details in the same tabbed panels (P2), and its visual encoding was cluttered with color schemes for performance and island membership.
Participants converged on filtering as a solution: P1, P5, P7, and P8 all suggested dimming nodes below a performance threshold.
P4 and P3 suggested novelty metrics to highlight qualitatively different nodes, and P6 wanted \textit{``performance jumps''} to be visually salient.
The system should highlight nodes most likely to merit attention and allow experts to customize what counts as noteworthy.

\subsubsection{Problem 2: Lack of Human-in-the-Loop Interaction}

Participants consistently expressed a desire to intervene in evolution.

\textbf{DR5: Enable targeted expert intervention.}\label{dr:5}
Participants wanted to track nodes of interest and act on them, with P1 requesting bookmarking.
E1 had similarly developed a manual workaround by writing down node IDs on paper.
P7 wanted to \textit{``choose a subset of nodes to continue evolving''} and prevent others from evolving to save resources. P8 shared the same idea.
We presented one design probe that included a suggestion dialog for \gls{nl}-guided mutation and a multi-select panel for merging.
Reactions were largely positive: P1 found the \textit{``suggest''} feature promising but noted that the system should \textit{``provide recommendation options''} rather than requiring guidance from scratch.
P5 is skeptical about merging, questioning whether it might \textit{``go against the original intent of evolutionary computation,''} while P4 and P6 responded positively.
The system should allow experts to intervene, including guiding mutations, combining ideas from multiple nodes, and shaping which lineages continue.

\textbf{DR6: Support flexible evolution pacing.}\label{dr:6}
Participants did not want a binary switch between full automation and manual control.
P5 and P8 noted that the desired involvement depends on situations and problems.
P7 articulated a similar vision: \textit{``fully automatic generation of some nodes, but with checkpoints where you can intervene at any time.''}
This points to a need for flexible pacing: the system should support a spectrum of involvement levels, from passive monitoring through throttled observation to step-by-step control, with the ability to shift between modes as the situation demands.

All DRs map to the theoretical frameworks introduced in Section~\ref{sec:related-work}.
\DR{1}--\DR{4} address information overload by supporting the three levels of Endsley's \gls{sa}~\cite{endsley_toward_1995}: \DR{3}--\DR{4} support perception (what is happening in the population), \DR{1}--\DR{2} support comprehension (why nodes differ), and \DR{3} also supports projection (which directions are promising).
\DR{5}--\DR{6} address the lack of steering,  moving the system from Sheridan's~\cite{sheridan_telerobotics_1992} Level~9 toward Level~5.


\section{\toolname}
\label{sec:tool}

\begin{figure*}[t]
  \centering
  \includegraphics[width=0.9\linewidth]{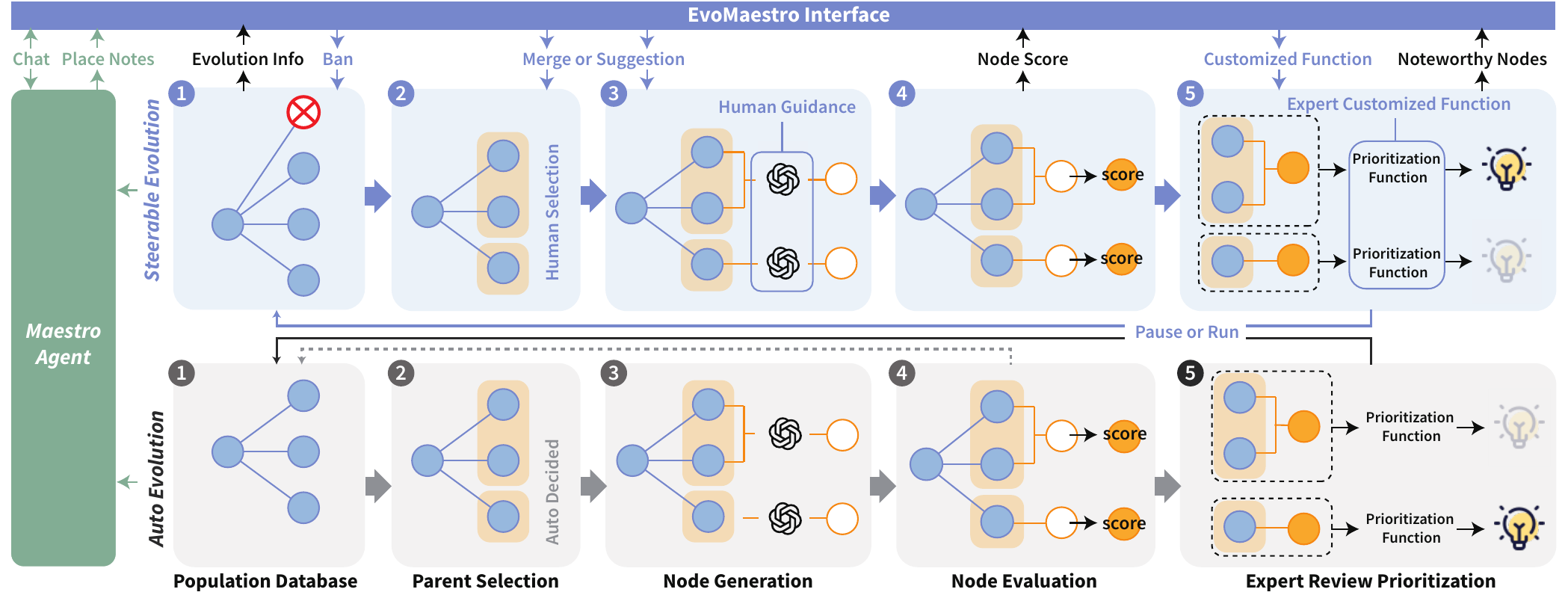}
  \caption{Steerable Program Evolution Framework.
    Stages \circled{1}--\circled{4} correspond to the standard loop (\S~\ref{sec:background});
    \circled{5} is added by \toolname.
    The steerable loop (Top) extends the automatic loop (Bottom) with Ban, Merge, Suggest, \crrev{Expert Review Prioritization}, and Pacing Control.
  The Maestro Agent (left) provides conversational code explanation and annotation.}
  \Description{A two-row conceptual diagram illustrating the steerable program evolution framework.
    At the top, a horizontal bar labeled EvoMaestro Interface lists the expert's input and output channels: Chat, Place Notes, Evolution Info, Ban, Merge or Suggestion, Node Score, Customized Function, and Noteworthy Nodes, with arrows connecting each channel to the corresponding loop stage below.
    The top row, labeled Steerable Evolution, shows five numbered stages flowing left to right.
    Stage 1 (Population Database) depicts a tree of blue circles with one node marked by a red ban symbol.
    Stage 2 (Parent Selection) shows a subset of the tree highlighted and labeled Human Selection.
    Stage 3 (Node Generation) shows two LLM icons, each producing an orange child node from a blue parent, with a Human Guidance label indicating that the expert can provide natural-language merge or suggestion instructions.
    Stage 4 (Node Evaluation) shows the generated nodes receiving score labels.
    Stage 5 (Expert Review Prioritization) shows evaluated nodes entering rectangular boxes labeled Prioritization Function, followed by lightbulb icons that indicate noteworthy outputs.
    The upper function is labeled Expert Customized Function, while the lower function represents the default prioritization.
    Upward arrows from Stages 4 and 5 feed Node Score and Noteworthy Nodes back to the interface.
    A Pause or Run control connects Stage 5 back to Stage 1.
    The bottom row, labeled Auto Evolution, mirrors the same five stages but without human intervention: parent selection is Auto Decided, generation uses no human guidance, and review prioritization uses the default function.
    A vertical green bar on the far left labeled Maestro Agent spans both rows, with arrows indicating that it reads from the Population Database and supports Chat and Place Notes.
  The overall layout emphasizes that the steerable mode adds human decision points to the otherwise automatic loop.}
  \label{fig:framework}
\end{figure*}

Guided by the six design requirements (\S~\ref{sec:formative-study}), we propose a framework for steerable program evolution and \toolname, an interactive visual analytics system built upon this framework (Fig.~\ref{fig:framework}).
The framework (\S~\ref{sec:tool-program-evolution-loop}) extends the standard evolution loop with intervention points and an expert review prioritization stage, enabling steering.
The interface (\S~\ref{sec:tool-interface}) organizes information through progressive disclosure to support understanding of an evolution at different levels of detail and provides views for steering the evolution.

\subsection{Steerable Program Evolution}
\label{sec:tool-program-evolution-loop}

This subsection describes how our framework transforms the fully automatic four-stage loop (\circled{1}--\circled{4}, \S~\ref{sec:background}) into a steerable process.
We first describe the four actions experts can take to intervene in a running evolution (\DR{5}, \DR{6}), then explain how these actions are enabled by our modifications to the loop stages.

When the expert identifies a node worth refining, they can use  \featurebf{Suggest} to select it as the parent node and provide \gls{nl} guidance to steer the next mutation.
The system feeds this guidance into the \gls{llm}'s prompt, producing a new child node.
To help the expert formulate their guidance, the system also generates a recommendation that the expert can adopt, modify, or ignore.
Additionally, the expert can use \featurebf{Merge} to select two nodes and direct the system to combine them into a new node with optional \gls{nl} instructions.
The system recommends merge candidates based on performance and dissimilarity (details in Appendix~\ref{sec:appendix-merge-recommendation}).
If the expert identifies a node pursuing an unproductive strategy, they can use \featurebf{Ban} to remove it from future parent selection, thereby pruning unproductive lineages.
For the entire loop, the expert can flexibly control the pace, pausing, stepping through one node at a time, or letting the evolution run automatically.
To enable these actions, our framework extends the standard four-stage loop at key stages (Fig.~\ref{fig:framework}, Top):

\textbf{\circled{1} Population Database.}
The expert can \feature{ban} nodes in the population database, preventing them from being selected in stage~\circled{2}.
Outside the loop, \feature{Maestro Agent}, an \gls{llm} assistant, reads from the population database to help experts understand code, compare nodes, and identify patterns across the evolution (\S~\ref{sec:sd-sensemaking}).

\textbf{\circled{2}--\circled{3} Parent Selection \& Node Generation.}
These stages are automatic in the standard loop.
Ours adds intervention points that allow the expert to override these stages through \feature{Suggest} or \feature{Merge}.

\textbf{\circled{4} Node Evaluation.}
At this stage, besides running evaluations, existing systems compute embeddings from each node's raw code for similarity comparison (\S~\ref{sec:background}).
\toolname instead computes \emph{reasoning-based embeddings} from each node's patch description and reasoning chain pair, capturing the similarity of \emph{algorithmic intent} rather than syntactic structure.
These embeddings underpin the clustering, dissimilarity chords, and comparison views throughout the interface.
Experts can toggle to code-based embeddings when they want to compare implementations instead of strategies.

\textbf{\circled{5} \crrev{Expert Review Prioritization} (new stage).}
The framework adds a fifth stage after node evaluation.
\crrev{A prioritization function consumes scores and embeddings from Node Evaluation and flags noteworthy nodes for inspection.
  By default, the function flags nodes whose evaluation score improves by at least 15\% relative to their parent.
  Experts can adjust the threshold or replace the function with one based on other signals, such as embedding dissimilarity.
Unlike ShinkaEvolve's novelty-based rejection sampling~\cite{lange_shinkaevolve_2025}, these flags affect only interface highlighting and do not alter scores, reject candidates, or affect parent selection.}

\textbf{Pacing control.}
The expert can switch at any time between three modes: Auto (free-running), Wait (pauses between generations), and Manual (one node at a time).

The steerable evolution loop provides the mechanisms for experts to intervene, while the \toolname interface surfaces these steering capabilities alongside features that support understanding the evolution at different levels of detail, which we describe next.

\subsection{\toolname Interface}
\label{sec:tool-interface}

The \toolname interface (Fig.~\ref{fig:interface}) is designed around a progressive workflow: the expert starts by gaining a global overview of the evolving population (\S\ref{sec:sd-overview}), then inspects specific nodes to understand their strategies and compare them (\S\ref{sec:sd-sensemaking}), and finally acts on those insights by steering the evolution (\S\ref{sec:sd-steering}).

\subsubsection{Evolution Understanding (\DR{3}, \DR{4})}
\label{sec:sd-overview}

Before drilling into individual nodes, the expert typically needs to \emph{get oriented} about the evolving population: what strategies have been explored, which nodes are performing well, and where the search might be heading.
They can then \emph{analyze the population structure} to identify clusters of similar strategies and the overall score distribution.

\textbf{Getting oriented.}
When the expert opens \toolname, a \emph{radial tree} in \feature{Population Overview} (Fig.~\ref{fig:interface}\captionID{A1}) provides an immediate visual summary of the full evolution history.
Nodes are arranged in concentric rings by generation and grouped into angular sectors by island (\captionID{A2}), with a sequential color scale encoding performance so that high-scoring nodes stand out at a glance.
Solid lines connect parent and child nodes, tracing evolutionary lineages; cross-island links indicate migrations where a node's offspring was assigned to a different island.
\crrev{We provide design rationale details in Appendix~\ref{sec:appendix-radial-layout}.}
To guide their focus, the best node in each island is marked with a golden circle \raisebox{-0.4ex}{\includegraphics[height=1em]{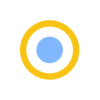}} (\captionID{A3}), and \crrev{noteworthy nodes prioritized for expert review} (\S~\ref{sec:tool-program-evolution-loop}) are flagged with a lightbulb \raisebox{-0.4ex}{\includegraphics[height=1em]{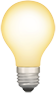}} (\captionID{A4}).
To understand how different the top strategies are from each other, they can read the \feature{Dissimilarity Chords} (\captionID{A5}), color-coded arcs connecting the best node of each island.
For a narrative summary without inspecting individual nodes, a sidebar panel (\captionID{A6}) provides an \gls{llm}-generated \feature{Evolution Overview} describing explored, promising, and failed strategies.
To reduce clutter, they can click on the encoded averages (the global one in the legend, per-island ones on seed nodes, and per-generation ones on arcs) to dim all nodes below that average; error and timeout nodes can be toggled off.

\textbf{Analyzing population structure.}
Two additional views help the expert understand patterns across the population.
\feature{Clusters View} (Fig.~\ref{fig:interface}\captionID{A7}) projects nodes into a scatter plot using PCA on their reasoning-based embeddings (\S~\ref{sec:tool-program-evolution-loop}), revealing clusters of nodes that share similar strategies.
\feature{Score Distribution} (\captionID{A8}) shows a density plot of scores; the expert can set a threshold to filter out low-scoring nodes, which are then dimmed on the tree.

\subsubsection{Node Understanding (\DR{1}, \DR{2})}
\label{sec:sd-sensemaking}

Beyond the population level, experts need to understand individual nodes.
\toolname supports this progressively: \emph{inspecting nodes} via hover tooltips and lineage paths for quick triage, \emph{understanding a node's strategy} through the detail panel and Maestro Chat with line-level annotations, and \emph{comparing nodes} through dissimilarity chords and side-by-side views.

\textbf{Inspecting nodes.}
Hovering over a node on the tree reveals a brief tooltip (Fig.~\ref{fig:interface}\captionID{B1}) showing its name, score, patch type, and island assignment, allowing the expert to quickly triage nodes without leaving the tree view.
Clicking a node selects it and draws an orange lineage path from the initial seed to that node, letting the expert trace its evolutionary history at a glance.
The last segment of this path uses a solid/dashed encoding: the solid half is drawn closer to the higher-scoring node of the pair, instantly revealing whether performance improved or regressed.

\textbf{Understanding a node's strategy.}
The detail panel (Fig.~\ref{fig:interface}\captionID{B2}) shows the node's performance scores, patch description, and metadata such as island assignment, node ID, and parent lineage.
The patch description provides a high-level understanding of the \emph{idea} that the patch pursued without requiring the expert to read the code.
For a deeper understanding, they can open the full source code (Fig.~\ref{fig:interface}\captionID{B3}) and invoke \feature{Maestro Agent} through the \feature{Maestro Chat} to explain it.
\feature{Maestro Chat} operates in two modes: in \feature{Code view}, the expert asks the agent to explain a single node's algorithmic strategy; in \feature{Code diff}, they view two nodes side by side and ask the agent to compare them.
Additionally, \feature{Maestro Agent} can place \textit{line-level annotations} directly on the code regions and highlight them.
This anchors abstract strategy descriptions to concrete code locations, which is more effective for supporting human judgment~\cite{IntelligentReasoningCues}.

\textbf{Comparing nodes.}
\feature{Dissimilarity Chords} (Fig.~\ref{fig:interface}\captionID{A5}) helps compare the best node in each island.
Experts can select multiple nodes for detailed comparison in \feature{Comparison Panel} (Fig.~\ref{fig:interface}\captionID{B4}), which shows performance comparisons and a dissimilarity heatmap.
They can also open a side-by-side code diff for any pair, with \feature{Maestro Chat} available to explain the differences.
The dissimilarity values throughout the interface are computed from reasoning-based embeddings (\S~\ref{sec:tool-program-evolution-loop}) for comparing algorithmic strategies.

\subsubsection{Evolution Steering (\DR{5}, \DR{6})}
\label{sec:sd-steering}

Having understood the population and identified promising or problematic nodes, the expert can act on those insights by guiding and combining nodes through \feature{Suggest} and \feature{Merge}, pruning unproductive lineages through \feature{Ban}, and controlling the pace of the evolution.
The steering mechanisms described in Section~\ref{sec:tool-program-evolution-loop} are accessed through a unified context menu (Fig.~\ref{fig:interface}\captionID
{C1}) that appears when right-clcking any node, along with \feature{Mark} and \feature{Note} for bookmarking and annotating nodes.

\textbf{Guiding, combining, and pruning nodes.}
Selecting \feature{Suggest} opens a modal (Fig.~\ref{fig:interface}\captionID{C}, top) where the expert writes \gls{nl} guidance, with an AI-generated recommendation as a starting point.
Selecting \feature{Merge} opens a modal (Fig.~\ref{fig:interface}\captionID{C}, middle) listing recommended merge partners ranked by score and dissimilarity.
Both actions produce new child nodes that appear on the tree in real time.
Selecting \feature{Ban} (Fig.~\ref{fig:interface}\captionID{C}, bottom) marks the node as banned, visually dimming it on the tree and removing it from future parent selection.

\textbf{Controlling the pace.}
\feature{Run Control} (Fig.~\ref{fig:interface}\captionID{C2}) allows switching between modes (\S~\ref{sec:tool-program-evolution-loop}), in which new nodes appear with a dashed border during evaluation, transitioning to solid once complete.

\section{System Demonstration}
\label{sec:case-study}

\begin{figure*}[t]
  \centering
  \includegraphics[width=\linewidth]{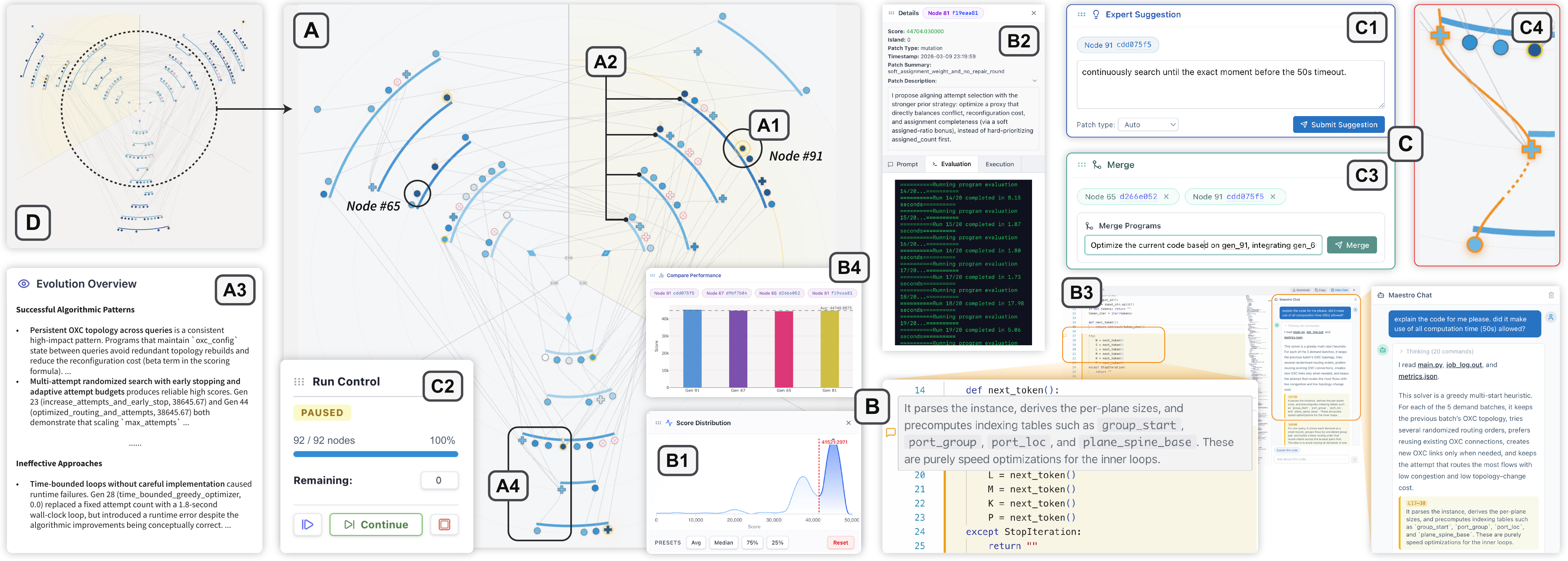}
  \caption{The expert's workflow during the system demonstration. (A)~Understanding: \feature{Population Overview} shows the evolution tree with best nodes highlighted (A1), per-generation score trends on arcs (A2), \feature{Evolution Overview} summarizing strategies (A3), and a banned node (A4). (B)~Inspecting nodes: \feature{Score Distribution} filters low-performing nodes (B1); the detail panel shows evaluation runtimes (B2); Maestro Chat explains node code with line-level annotations (B3); and \feature{Comparison} reveals dissimilarity among top nodes (B4). (C)~Steering: \feature{Suggest} injects time-budgeting guidance (C1); \feature{Run Control} sets the node budget (C2); and \feature{Merge} combines complementary nodes (C3). (D)~The final evolution tree with the best node highlighted.}
  \Description{A composite figure with annotated screenshots arranged in four groups labeled A through D, showing the expert's workflow during the OXC routing system demonstration. Group A (Understanding) contains four panels: A1 shows the Population Overview radial tree with approximately 90 nodes across multiple islands, where the best nodes are highlighted with orange fills and one node is marked with a red ban symbol (A4); A2 shows small sparkline score-trend charts embedded along the tree arcs between generations, revealing per-generation performance trajectories; A3 shows the Evolution Overview panel listing successful algorithmic patterns (such as persistent OXC topology across queries and multi-attempt randomized search with early stopping) and ineffective approaches (such as time-bounded loops causing runtime failures) as bullet-pointed text. Group B (Inspecting Nodes) contains four panels: B1 shows the Score Distribution histogram with a cursor filtering nodes by score range, where most scores cluster between 10,000 and 50,000; B2 shows the Node Details panel for Node 81, displaying its score (64704.033000), island, patch type (mutation), timestamp, patch summary, and patch description text proposing to align attempt selection with a stronger prior strategy, along with Prompt, Evaluation, and Execution tabs where the Evaluation tab is active showing timestamped log lines of the evaluation process running 14 out of 28 test cases; B3 shows a code editor displaying Python source code with syntax highlighting for a function called next\_token, with inline Maestro Agent annotations explaining that the code parses the instance, derives per-plane sizes, and precomputes indexing tables, and a Maestro Chat panel on the right where the expert asks whether the code makes use of all computation time and the agent responds with a detailed explanation of the solver's greedy multi-start heuristic; B4 shows a Compare Performance panel with four selected nodes listed by ID, displaying a grouped bar chart comparing their scores across multiple evaluation dimensions with colored bars. Group C (Steering) contains four panels: C1 shows an Expert Suggestion modal targeting Node 91, with the text instruction to continuously search until the exact moment before the 50-second timeout, a patch type selector set to Auto, and a Submit Suggestion button; C2 shows the Run Control panel in PAUSED state at 92 of 92 nodes (100 percent) with Remaining set to 0 and Continue and stop buttons; C3 shows a Merge modal with two selected nodes (Node 65 and Node 91) and a text field containing the instruction to optimize the current code based on gen\_91 integrating gen\_6, with a Merge button; C4 shows the EvoMaestro interface icon with a blue plus symbol indicating a new node was created. Group D shows the final evolution tree after steering, now containing over 200 nodes with the best-performing node highlighted, demonstrating substantially more nodes than the initial tree in panel A1.}
  \label{fig:case-study}
\end{figure*}





\crrev{To assess \toolname in a complex real-world setting, we present a system demonstration with E1, a co-author and domain expert who contributed feedback during the formative phase.}
\crrev{E1 was particularly suited to this extended demonstration because he combined extensive experience with the baseline system, relevant algorithm-design expertise, strong motivation to apply \gls{llm}-driven program evolution in his research in theoretical computer science and algorithm design, and willingness to commit substantial time.}
We adapted an optimization problem from the ICPC 2025 Online Winter Challenge\footnote{\url{https://icpc.global/regionals/finder/Winter-Challenge-2026}} as the task. It aims to optimize traffic routing in a large-scale AI computing cluster interconnected by \glspl{oxc}. Appendix~\ref{sec:appendix-oxc} provides the task details in depth.
E1 used \toolname for approximately 16 hours over 7 days for the task. \crrev{The following describes his workflow, which we reconstructed from several types of data, including his contemporaneous activity log of important actions and reasoning, a short post-study summary, system logs of evolution actions, \feature{Maestro Chat} histories, and generated node, score, and lineage artifacts. During the reconstruction, we triangulated his rationale with recorded actions and generated artifacts, and followed up with E1 when details were unclear. We did not log every low-level UI action (\eg{} tab switch or node click).}

To kickstart the evolution process, E1 generated an initial seed program using GPT-5. During the study, E1 interacted with \toolname through a recurring analysis loop: (1) understanding the current evolution process, (2) inspecting specific nodes for steering, (3) intervening to inject domain knowledge, and (4) setting the system to evolve automatically before re-observing the results.
We illustrate each step of this loop with concrete examples (Fig.~\ref{fig:case-study}).

\textbf{Understanding the current evolution process.}
At the beginning of each cycle, E1 typically sought a global overview, including best nodes and overall evolutionary trend.
For example, after the initial 92 nodes were generated automatically, he first located the best nodes (both global and island-level) in \feature{Population Overview} (Fig.~\ref{fig:case-study}\captionID{A}), highlighted by golden circles \raisebox{-0.4ex}{\includegraphics[height=1em]{figures/4-system-design/best-node.png}} (Fig.~\ref{fig:case-study}\captionID{A1}), then skimmed the arcs (Fig.~\ref{fig:case-study}\captionID{A2}) to preview the evolutionary trend.
After confirming that the evolutionary trend was generally improving (\ie{} arc colors became deeper across generations), he examined the \feature{Evolution Overview} (Fig.~\ref{fig:case-study}\captionID{A3}) to obtain a strategy summary.
It showed that the top-performing nodes had converged on two strategies (multi-attempt randomized greedy search and structural deterministic routing) and that earlier time-bounded search loops had largely failed, whereas successful nodes used adaptive attempt budgets.
This raised a key question: were the top-performing nodes fully utilizing the 50-second execution time limit? To answer this, he proceeded to inspect specific nodes.

\textbf{Inspecting nodes for steering.}
When something at the global level caught his attention, E1 inspected individual nodes to verify hypotheses or seek explanations.
To investigate the time-budgeting question, he first used the \feature{Score Distribution} (Fig.~\ref{fig:case-study}\captionID{B1}) to filter out low-performing nodes and focus on top candidates. He then examined the detail panels of selected nodes (N-54, 67, 75, 81, and 91, the best node) and reviewed their evaluation results across test cases (Fig.~\ref{fig:case-study}\captionID{B2}).
The longest runtime was 20s, far below the limit, confirming that the node was leaving significant computation time unused.
He then used \feature{Maestro Chat} (Fig.~\ref{fig:case-study}\captionID{B3}) with line-level annotations to understand the best node's search logic in detail.
Surprised by the sophisticated algorithm (precomputation, pruning, and steal mechanisms) implemented in the best node (N-91), he further explored other nodes (N-65, 67, 81), guided by the noteworthy-node flags \raisebox{-0.4ex}{\includegraphics[height=1em]{figures/4-system-design/lightbulb.png}}.
He compared their performance and dissimilarity and found that although they have similar scores, N-91 is distinct from the other three nodes, as shown in \feature{Comparison Panel} (Fig.~\ref{fig:case-study}\captionID{B4}).
He further used \feature{Maestro Chat} to understand their code details.

\textbf{Intervening with domain knowledge.}
Once E1 identified an actionable insight during node inspection, he intervened in the evolution process.
Continuing the above example, he used \feature{Suggest} (Fig.~\ref{fig:case-study}\captionID{C1}) on N-91 (44k) to inject time-budgeting knowledge: \textit{``continuously search until the exact moment before the 50s timeout,''} producing N-94 (44.5k).
He then generated 4 new nodes (N-95 - N-98) using \feature{Run Control} (Fig.~\ref{fig:case-study}\captionID{C2}).
Based on insights from node inspection, he found the deterministic waterfilling algorithm in N-65 complementary to N-91, and thus used \feature{Merge} (Fig.~\ref{fig:case-study}\captionID{C3}) to combine them, producing N-99 (45.0k).
Beyond these targeted interventions, E1 also explored speculative ideas using \feature{Suggest}, adapting his strategies based on observed patterns.
For example, by inspecting improvement trends through arc and node colors, he observed that one lineage (N-26->29->48) lagged behind the overall generational progress (Fig.~\ref{fig:case-study}\captionID{A4} and \captionID{C4}). After discussing this in \feature{Maestro Chat}, he concluded that this lineage introduced excessive complexity with only marginal gains, and used \feature{Ban} to remove N-48.

\textbf{Resuming automatic evolution.}
After each round of intervention, E1 resumed automatic evolution.
Before resuming, he typically used \feature{Ban} to prune unproductive nodes, preventing the system from revisiting dead ends.
He then set a node budget in \feature{Run Control} (\eg{} 20 new nodes) and left to do other work.
Upon returning, he examined the \feature{Population Overview}, \feature{Evolution Overview}, and \feature{Score Distribution} to assess whether the injected knowledge had taken effect, thereby initiating the next analysis cycle.

This cycle was repeated over the 7-day period. E1 progressively shifted from active steering to passive monitoring as the evolution matured.
By the end, the system had generated over 200 nodes with the best score of 45.8k (N-144) across multiple distinct strategies (Fig.~\ref{fig:case-study}\captionID{D}).
Throughout the process, he relied entirely on observation and steering through \toolname, without writing code.
In his post-study summary, E1 reported that \toolname shifted his work from tedious code review toward active human-AI collaboration. He identified \emph{Merge}, \emph{Suggest}, and \emph{Maestro Chat} as central to this shift and expressed interest in using \toolname in future research.


\section{User Study}
\label{sec:evaluation}

We conducted a within-subjects user study to compare \toolname against the baseline.
Our goals were to assess whether \toolname improves experts' ability to (1)~understand the state and strategies of an evolving population, (2)~identify decision-relevant information, and (3)~intervene effectively in an evolution.
\crrev{Accordingly, this comparison evaluates understanding, workload, and steering support, not final program quality.}
We describe the study setup first and then report the results.

\subsection{Study Setup}
\label{sec:eval-method}

With the same recruitment process as the formative study (\S~\ref{sec:fs-method}), we recruited \usnum{} participants (\textbf{U1--U12}; 4 female, 8 male). Three (U3, U6, U9) were also participants in the formative study.
All participants had experience in algorithm design or optimization, including software engineers (U3, U9) and graduate students (U1, U2, U4--U8, U10--U12) spanning \gls{dl}, physics, statistics, \gls{hci}, and mechanical engineering.
All are fluent in Python and daily users of \gls{llm}-based programming tools.
Most had basic familiarity with \gls{llm}-based evolutionary systems; two (U2, U11) had no prior exposure and one (U8) had long-term hands-on experience with similar systems.
Participant details are provided in Table~\ref{tab:eval-participants} (Appendix~\ref{sec:appendix-eval-participants}).

\textbf{Study design:}
We used a within-subjects design comparing \toolname against the baseline ShinkaEvolve~\cite{lange_shinkaevolve_2025} web UI.
Each participant used both conditions on two multi-objective optimization tasks of comparable complexity (details in Appendix~\ref{sec:appendix-eval-tasks}), counterbalanced using a $2 \times 2$ Latin square (Appendix~\ref{sec:appendix-eval-counterbalancing}).
In the \toolname condition, the evolution remained live so that participants could execute real-time interventions.

\textbf{Procedure:}
Each 80-minute session comprised a tutorial on \gls{llm}-driven evolution, followed by two 25-minute task sessions (one per condition).
In each task session, participants first spent 15 minutes on sensemaking tasks: identifying the most promising nodes, describing algorithmic strategies, and spotting evaluation weaknesses.
They then spent 10 minutes on intervention tasks: in the \toolname condition, writing a \gls{nl} suggestion and initiating a merge; in the baseline condition, describing their intended interventions in writing, capturing intervention \emph{intent} for qualitative comparison.
After each condition, participants completed a questionnaire and NASA-TLX.
A post-study interview compared experiences across both conditions.
The details are provided in Appendix~\ref{sec:appendix-eval-procedure}.

\textbf{Measures:}
We designed a custom 7-point Likert questionnaire to assess three aspects mapped to our design requirements.
\emph{Understandability} (5 items, both conditions) measured whether each system helped participants understand the population's evolution, discern strategies, locate notable nodes, compare nodes, and identify individual node algorithms, covering \DR{1}--\DR{4}.
\emph{Usability} (3 items, both conditions) assessed ease of use, information organization, and task effectiveness.
These shared items enable paired statistical comparison between conditions.
Three additional \emph{Intervenability} items (\toolname only) assessed intervention identification (\DR{5}), pacing control (\DR{6}), and intervention effectiveness (\DR{5}), and one item assessed the value of \emph{Maestro Chat} for code understanding.
These \toolname-only items capture capabilities with no baseline equivalent.
We also administered the NASA-TLX~\cite{hart_development_1988} after each condition to measure cognitive workload and collected open-ended responses and a post-study interview for qualitative analysis.
Full questionnaire items are provided in Appendix~\ref{sec:appendix-eval-measures}.
For shared items, we use the Wilcoxon signed-rank test with Cliff's delta ($\delta$) for effect sizes~\cite{cliff_dominance_1993} where $|\delta|<0.33$ is small, $<0.474$ medium, and $\geq0.474$ large. \toolname-only items are reported descriptively.
Qualitative analysis followed the same approach in the formative study (\S~\ref{sec:fs-analysis}).


\subsection{Results}
\label{sec:eval-results}

We report the results from our Likert questionnaires and NASA-TLX, followed by qualitative findings from the open-ended questions and post-study interviews.

\subsubsection{Closed-ended Questions}

Overall, \toolname significantly outperformed the baseline on all eight shared Likert items and received consistently high ratings on the \toolname-only items.

\begin{figure*}[t]
  \centering
  \includegraphics[width=\linewidth]{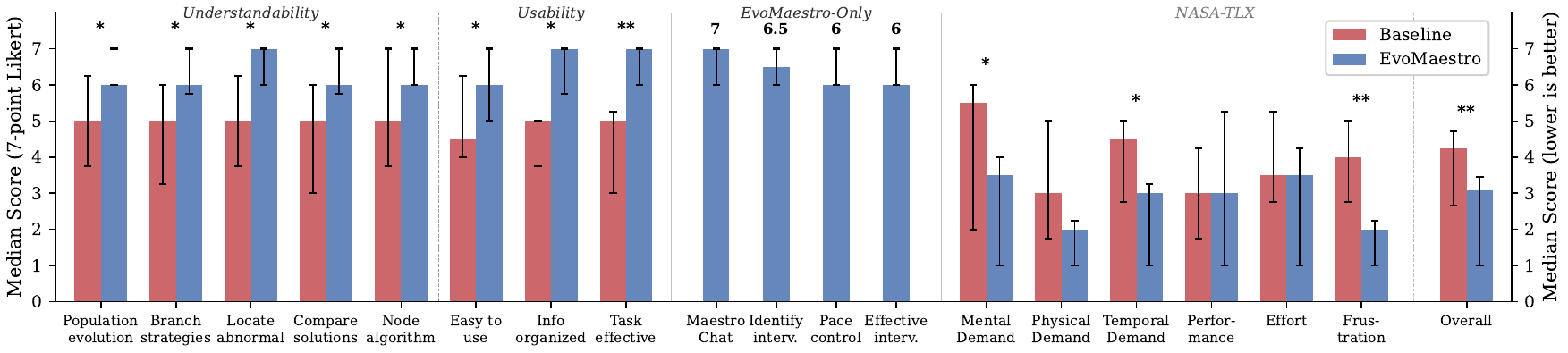}
  \caption{Evaluation results (detailed values in Tables~\ref{tab:likert-results} and~\ref{tab:tlx-results} in Appendix~\ref{sec:appendix-eval-results}). Bars show median scores with interquartile ranges (Q1–Q3). Left: Shared Likert items (7-point scale; higher is better). Middle: \toolname-only Likert items (numbers indicate medians).
    Right: NASA-TLX scores (lower is better); \crrev{its Performance subscale is participants' perceived success in completing the task.}
  Asterisks mark significant differences.}
  \Description{A horizontal bar chart divided into four sections by vertical dashed lines, with the left y-axis labeled Median Score on a 7-point Likert scale (higher is better) and the right y-axis labeled Median Score (lower is better) for the NASA-TLX section. Each item shows two side-by-side bars: red for Baseline and blue for EvoMaestro, with black error bars spanning the interquartile range from Q1 to Q3. Asterisks above bar pairs indicate statistical significance (* for p less than 0.05, ** for p less than 0.01). In the Understandability section (five items): Population evolution shows Baseline median 5 versus EvoMaestro median 6 (*); Branch strategies shows 5 versus 6 (*); Locate abnormal shows 5 versus 7 (*); Compare solutions shows 5 versus 6 (*); Node algorithm shows 5 versus 6 (*). In the Usability section (three items): Easy to use shows 4.5 versus 6 (*); Info organized shows 5 versus 7 (*); Task effective shows 5 versus 7 (**). In the EvoMaestro-Only section (four items, blue bars only with median values printed above): Maestro Chat median 7; Identify intervention median 6.5; Pace control median 6; Effective intervention median 6. In the NASA-TLX section (seven items, using the right y-axis where lower is better): Mental Demand shows Baseline median 5.5 versus EvoMaestro median 3.5 (*); Physical Demand shows 3 versus 2 (not significant); Temporal Demand shows 4.5 versus 3 (*); Performance shows 3 versus 3 (not significant); Effort shows 3.5 versus 3.5 (not significant); Frustration shows 4 versus 2 (**); Overall shows approximately 4.25 versus 3.08 (**). All error bars for both conditions span roughly 2 to 3 points on the scale.}
  \label{fig:eval-results}
\end{figure*}

\textbf{Understandability \& Usability.}
Fig.~\ref{fig:eval-results} (left) summarizes the comparison on shared Likert items.
\toolname scored significantly higher than the baseline on all eight shared items.
The strongest are task effectiveness ($\delta=0.694$, large) and information organization ($\delta=0.639$, large), followed by locating abnormal nodes (\DR{4}: $\delta=0.521$, large) and comparing nodes (\DR{2}: $\delta=0.486$, large).
Overall, the participants rated \toolname's understandability at a median of 6.0 (mean 6.20) versus 5.0 (mean 4.65) for the baseline, and usability at 7.0 (mean 6.31) versus 5.0 (mean 4.61).
The largest improvements were in usability items, suggesting that our progressive disclosure and visual encodings make the evolution process not only more comprehensible but also much easier to work with.

\textbf{\toolname-only items.}
The Intervenability and \feature{Maestro Chat} items were consistently rated highly (Fig.~\ref{fig:eval-results}, middle).
AI-assisted code explanation (\feature{Maestro Chat}) received a median of 7.0 (mean 6.42, SD 0.90), the highest rating of any individual item, reflecting participants' strong appreciation for \gls{llm}-mediated code understanding.
The three intervenability items received medians of 6.0--6.5, indicating that participants found the steering mechanisms effective.
\toolname's design helps experts see \emph{where} to intervene, while the act of formulating effective guidance remains more challenging, as identifying intervention needs (\DR{5}, median 6.5) scored slightly higher than conducting interventions (\DR{5}, median 6.0).

\textbf{Cognitive workload (NASA-TLX).}
Fig.~\ref{fig:eval-results} (right) summarizes the NASA-TLX results.
\toolname significantly reduced overall cognitive workload (the mean of all six subscales per participant) compared to the baseline (median 3.08 vs.\ 4.25; $p=0.004$, $\delta=-0.417$, medium effect).
For the subscales, the largest improvements were in \emph{frustration} ($p=0.004$, $\delta=-0.535$, large), \emph{temporal demand} ($p=0.016$, $\delta=-0.479$, large), and \emph{mental demand} ($p=0.043$, $\delta=-0.347$, medium).
The frustration reduction is the largest effect, which we attribute to the combination of progressive disclosure design (reducing information overload) and steering mechanisms (enabling experts to act on their observations rather than passively watching).
Both systems enabled participants to complete the tasks, as the Performance and Effort subscales did not differ, but \toolname did so with much less cognitive strain.

\subsubsection{Qualitative Findings}

We organize the qualitative findings from post-study interviews and open-ended responses around three concerns: understanding, steering, and usability.

\textbf{Evolution Understanding.}
The feedback has a consistent contrast between the two systems regarding information structuring.
The baseline presented all data at once, which overwhelmed users.
U4 and U9 reported confusion and cognitive overload due to the large number of nodes and complex visualizations.
\toolname's progressive disclosure approach relieves users of information overload. U5 noted a significantly reduced mental load, describing the experience as \textit{``peeling away layers.''}
U4 found the radial tree \textit{``much clearer than ShinkaEvolve's style of piling everything together.''}
These findings suggest that the value of our design lies not in showing less information but in revealing it in the right order: population-level orientation first, then node-level details on demand.

\textbf{Evolution Steering.} All participants appreciated the ability to intervene. U5 noted that \textit{``we can add constraints to the LLM, and this is visible on the system, making it easy to trust,''} and U10 noted that \feature{Ban} resolved their concern about \textit{``the AI wasting tokens repeating the same strategy.''}
Despite this appreciation, their comments revealed a nuanced relationship between automation and human control.
Rather than replacing automation, participants used \gls{llm}-generated recommendations as a starting point: U7 described \textit{``checking the merge recommendations, examining the recommended nodes in detail, then finding further inspiration,''} and U4 noted that the recommendations \textit{``deepened my understanding of the algorithms,''} suggesting they served as learning aids rather than just shortcuts.
Looking forward, participants wanted even richer automation support, such as \textit{``global recommended actions''} (U9) and guiding groups of related nodes simultaneously (U7, U8).

\textbf{Overall Usability.}
\gls{nl} guidance is crucial for improving the usability of such complex systems with dense information, as \feature{Maestro Chat} was the highest-rated feature (median 7.0).
U5 noted its usefulness at different levels: \textit{``not only can it provide global understanding, but also understanding of individual lines or functional blocks, saving a tremendous amount of time.''}
However, the system's usability did not rely on \feature{Maestro Chat} alone.
U4, who had competitive programming experience but a limited optimization background, chose not to use it because of its verbose responses but still found the system accessible through interactive steering: \textit{``while playing around with Merge, I actually came to understand a bit of what optimization researchers do.''}
The system offers interactions that suit users with  different backgrounds:
experts with strong algorithmic knowledge can leverage \feature{Maestro Chat} for deep code understanding, while those newer to the problem domain can build their understanding through hands-on interaction with the steering features.


\section{Discussion}
\label{sec:discussion}


\crrev{Designing and evaluating \toolname highlights the importance of expert agency in open-ended \gls{llm}-driven search.}
We discuss how understanding and steering support this agency, then examine current limitations and future directions.

\subsection{\texorpdfstring{\crrev{Lessons Learned}}{Lessons Learned}}

\crrev{\Gls{llm}-driven program evolution creates two linked challenges for expert agency.
  A growing population can exceed what an expert can inspect linearly, limiting their ability to judge what the system is doing.
  Meanwhile, full automation leaves domain insights unused when experts cannot redirect the search.
Our experience suggests that preserving expert agency requires both forming an informed judgment and making that judgment affect subsequent automation.}

\crrev{\textbf{Understanding preserves expert agency.}
  Information overload cannot be addressed by simple information hiding or summarization.
  Scores identify relative performance, but do not explain how a candidate works or whether its strategy makes sense for the problem.
  Semantic oversight therefore follows three steps: (1) route attention using population structure, filters, noteworthy-node flags, and lineage; (2) compare algorithmic strategies at an appropriate level; and (3) inspect code, diff, logs, and evaluation results when verification is needed.
\feature{Maestro Chat} and its explanations help experts move between code and strategy, while the code and evaluation results remain available for checking.}
\crrev{In our evaluation, some of the strongest improvements concerned information organization, locating abnormal nodes, and comparing nodes, while overall cognitive workload decreased.}
\feature{Maestro Chat} was the highest-rated feature, suggesting that when \gls{llm} agents produce complex artifacts, dedicating another agent to explanation can substantially reduce the cognitive cost of human oversight.
\crrev{More broadly, by presenting supporting evidence in a useful order, such interfaces can help experts form their own judgment about a candidate and identify when closer inspection is needed.}

\crrev{\textbf{Steerability turns automation into collaboration with trust.}
  Forming a judgment is only one part of agency, because understanding alone leaves the expert just as an observer. Experts also need to act on it.
  Collaboration requires trust and a way for domain knowledge to change what the automated process does next.
In \toolname, the significance of its intervention features is not that selection, rejection, and recombination are new to \gls{ea}, but that experts can use these operations based on algorithmic ideas to change which ideas later generations develop.}
\crrev{The immediate effects of these interventions are visible in the population, so experts can inspect how the population responds before deciding whether to intervene again.}
Participants rated the intervention support highly. They consistently linked the ability to intervene with their trust in the system's output. U1 called interventions \textit{``necessary for producing better results and trusting those results,''} and U5 valued that the constraints added by the expert \textit{``are visible in the system, making it easy to trust.''}
\crrev{The system demonstration also illustrates how an expert moved between active intervention and passive monitoring.
The broader lesson is that collaboration does not require constant human control but trust and a path for expertise to affect automation when judgment is needed.}

\crrev{Together, these lessons connect agency to task-level alignment in \gls{llm}-driven search.
  Initial prompts and evaluators encode the task objective, but may not anticipate every failure mode or concern that emerges during open-ended search.
  Without understanding, experts cannot verify whether emerging strategies remain consistent with their intent; without steering, they cannot redirect the search when strategies diverge.
Task-level alignment should therefore be supported as an ongoing process of inspection, correction, and monitoring, rather than only through an initial specification.}

\subsection{Limitations and Future Directions}

We also acknowledge our limitations and discuss future directions.

\crrev{\textbf{System demonstration with a co-author.}
  E1 was a co-author who provided feedback during the formative phase.
  Although he had not seen the final \toolname system before the demonstration, his prior involvement gave him expectations about the system's capabilities.
  Together with his strong motivation and substantial time investment, this may have encouraged deeper and more persistent engagement than might be expected from other users and may have shaped his positive reflections.
  We therefore use the demonstration only to illustrate how sustained expert steering unfolded through his workflow.
Future case studies with independent domain experts are needed to examine whether these usage patterns generalize.}

\crrev{\textbf{Reliability of generated explanations.}
  \Gls{llm}-generated patch descriptions, summaries, chat responses and annotations may be plausible yet incorrect.
  \toolname allows for progressive in-depth inspection, but experts may still miss errors, especially outside their areas of expertise.
Future systems should expose stronger provenance and support cross-checking and validation across models.}

\crrev{\textbf{Scalability.}
  Our evaluation involved evolution trees with 50 nodes, and the system demonstration produced approximately 200 nodes.
  Filtering helped experts focus at these scales, but does not establish that the interface can support thousands of nodes.
Larger populations may require hierarchical aggregation, collapsed lineages, or focus-and-context techniques.}

Future work can examine how our approach can be extended or transferred to other domains and iterative multi-agent systems.

\textbf{Interaction extensions.}
Participants wanted to guide groups of related nodes, suggesting future support for recommending and steering node groups.
The system could also suggest pacing changes, such as automatic pausing when a population converges.

\crrev{\textbf{Beyond software engineering.}
  Programs can represent candidate solutions outside software engineering when they can be executed and evaluated with metrics (\S~\ref{sec:related-work}).
  Potential users may have deep domain knowledge without equivalent programming expertise or the time to inspect every generated program directly.
  The \gls{hci} challenge is therefore to connect domain concepts, algorithms, code, and evaluation results.
  Our studies cover algorithmic and numerical optimization, and interfaces for other fields may require different representations and interventions.
Evaluating the approach in these fields remains future work.}

\textbf{From co-author to curator: Beyond program evolution.}
Most \gls{hci} work on \gls{llm}-assisted programming studies a single developer working with a single assistant~\cite{vaithilingam_expectation_2022, barke_grounded_2023, mozannar_reading_2024}.
\crrev{However, as \gls{llm}-based multi-agent systems advance beyond program evolution, these lessons learned may also inform the design of other iterative multi-agent systems that maintain persistent revision histories.
Users in such systems similarly need to locate, compare, combine, and redirect alternatives generated along different trajectories. Evaluating this transfer remains future work.}

\section{Conclusion}
\label{sec:conclusion}


Existing \gls{llm}-driven program evolution systems operate as fully automated black boxes that provide limited support for domain experts to understand and steer an evolution.
Through a formative study with \fsnum{} domain experts, we identified two core challenges (understanding and steering) and derived six design requirements.
We then developed a steerable program evolution framework that enables expert intervention and \toolname, \crrev{which supports semantic oversight through multi-level inspection and interventions based on algorithmic ideas.}
A within-subjects study with \usnum{} participants showed significantly higher ratings across all shared understandability and usability items, high ratings for intervention support, and reduced cognitive workload.
\crrev{A system demonstration illustrated sustained expert steering across more than 200 nodes over 7 days.}
In future work, we plan to evaluate the approach with independent domain experts, larger populations, tasks in other fields, and stronger support for verifying generated explanations.
\crrev{More broadly, our work suggests that preserving expert agency in open-ended AI search requires both helping experts form informed judgments and allowing them to shape subsequent automation.}

\begin{acks}
  We used \glspl{llm} for proofreading and \LaTeX{} programming but wrote and edited the text content of this paper manually.
  This study is supported by the Ministry of Education, Singapore, under Academic Research Funds (NTU Tier 1, RG104/25 and Tier 2, Proposal ID: T2EP202220049). Any opinions, findings, conclusions, or recommendations expressed are those of the author(s) and do not reflect the views of the Ministry of Education, Singapore.
\end{acks}

\bibliographystyle{ACM-Reference-Format}
\bibliography{main}

\appendix

\section{Formative Study Details}
\label{sec:appendix-formative-study}

\subsection{Participants}
\label{sec:appendix-fs-participants}

\begin{table}[H]
  \caption{Formative study participants. \gls{ea} familiarity is self-reported during the interview: ``None'' indicates no prior exposure to \glspl{ea}; ``Basic'' indicates familiarity with core concepts (\eg{} population, mutation, selection) but no implementation experience. None had implemented or published \gls{ea}-related work. All participants are daily users of \gls{llm}-based programming tools.}
  \label{tab:fs-participants}
  \small
  \begin{tabular}{p{0.04\linewidth} p{0.25\linewidth} p{0.40\linewidth} p{0.12\linewidth}}
    \toprule
    \textbf{ID} & \textbf{Background} & \textbf{Domain} & \textbf{\gls{ea} Famil.} \\
    \midrule
    P1\textsuperscript{\dag} & Software Engineer & Database Optimization & Basic \\
    P2 & Software Engineer & Backend Optimization & None \\
    P3 & PhD Student & \gls{dl} & None \\
    P4\textsuperscript{\dag} & PhD Student & Mathematics and \gls{dl} & None \\
    P5\textsuperscript{\dag} & PhD Student & \gls{dl} & None \\
    P6\textsuperscript{\dag} & Software Engineer & Backend Optimization & Basic \\
    P7 & MSc Student & \gls{dl} & None \\
    P8 & MSc Student & \gls{dl} & None \\
    \bottomrule
  \end{tabular}

  \textsuperscript{\dag}Over 5 years of competitive programming experience.
\end{table}

\subsection{Procedure}
\label{sec:appendix-fs-procedure}

Each formative study session followed a seven-stage protocol.
Sessions were conducted with audio and screen interactions recorded with consent.

\textbf{Stage 1: Consent and warm-up (2 min).}
We obtained informed consent and established rapport.

\textbf{Stage 2: Background questions (8 min).}
We asked participants about their research domain, their experience iteratively improving algorithms with \gls{llm} tools, how they manage multiple nodes, and how they evaluate node quality.
These questions were designed to surface current workflow pain points before introducing the evolution concept.

\textbf{Stage 3: Concept introduction (5 min).}
We introduced \gls{llm}-driven program evolution using a three-part visual handout covering the evolutionary loop (population, selection, \gls{llm} mutation, evaluation), the distinction between random and reasoned mutations, and the island model for maintaining diversity.
We did not mention \toolname or any specific design ideas at this stage.

\textbf{Stage 4: System walkthrough and free exploration with think-aloud (15 min).}
After a brief orientation to the ShinkaEvolve baseline interface\footnote{Available at \url{https://shinka-mop.reify.ing}}, participants freely explored a pre-run evolution result containing approximately 50 nodes across multiple islands, generated for a drone path optimization task with competing objectives (flight efficiency, noise mitigation, and signal stability).
We asked participants to think aloud as they explored and used minimal prompts (\eg{} ``What are you looking at right now?'' ``What would you want to know that you can't see?'') only when they fell silent for more than ten seconds.

\textbf{Stage 5: Semi-structured follow-up (10 min).}
We asked targeted questions covering information usefulness, code comprehension difficulty, strategies for identifying promising nodes, understanding of the overall evolution trajectory, desire to intervene, opportunities for expert knowledge injection, and preferences for pacing and control.

\textbf{Stage 6: Design probes (8 min).}
We showed two design probes framed as ``ideas we are exploring'':
Probe~A demonstrated information overload reduction (score-emphasis tree with performance color coding, noteworthy-node highlighting, and per-node strategy summaries);
Probe~B demonstrated expert steering (a suggestion dialog for \gls{nl}-guided mutation and a merge panel for multi-parent recombination).
We collected reactions, suggestions, and concerns for each probe.

\textbf{Stage 7: Wrap-up (2 min).}
We asked participants to summarize in one sentence what they found most difficult about understanding the evolution run and invited any remaining thoughts.


\section{Design and System Details}
\label{sec:appendix-design-system-details}

\subsection{Radial Layout Design}
\label{sec:appendix-radial-layout}

\crrev{\textbf{Design rationale.}
  Evolution histories usually begin with a few seed nodes and expand as descendants accumulate.
  The radial tree places early generations near the center and later generations on larger circumferences, giving denser outer generations more space.
  In contrast, a top-down tree uses a rectangular band at each depth, leaving much of the band unused near sparse early generations while later branching histories compete for the same width.
  The radial layout also maps the evolution structure directly: radius encodes generation depth, angular sectors group islands, and edges trace lineage.
  \toolname also provides a sortable node table for exact score and metadata lookup, while the radial tree serves as the process view for generation, lineage, island context, and intervention history.
  Exact angular positions within a sector are layout-driven and carry no data meaning.
Euclidean distance is not designed to encode strategy similarity; only its radial component reflects generation difference.}

\begin{figure}[t]
  \centering
  \includegraphics[width=\columnwidth]{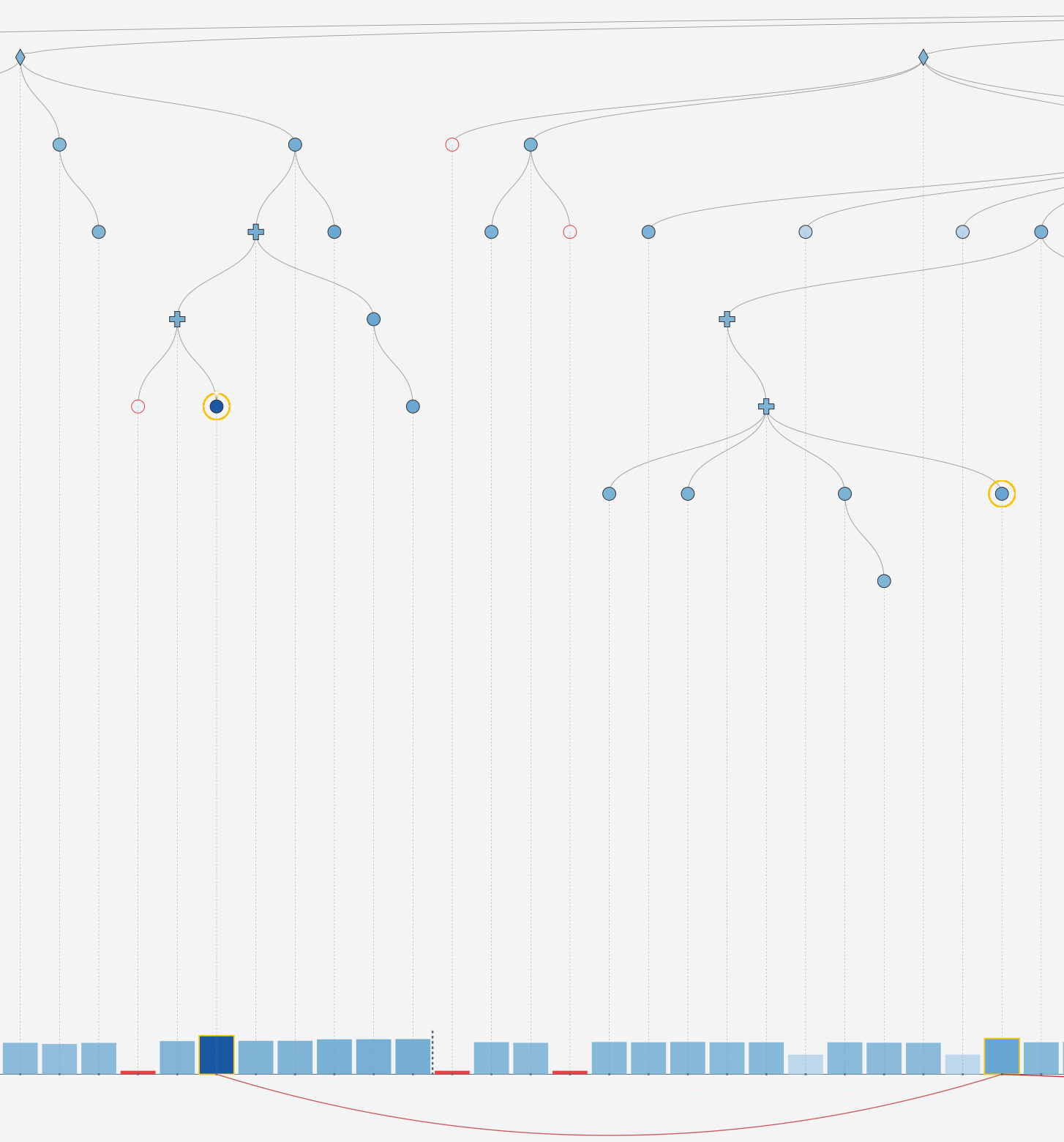}
  \caption{\crrev{Alternative tree-and-bar design considered during design iteration. Each node has a unique horizontal position, and a dotted guide links it to an aligned score bar below the tree. Arcs connect node pairs whose dissimilarity exceeds a threshold. As the population grew, the lineage tree, score bars, and arcs competed for horizontal space.}}
  \Description{A top-down evolution tree with each node vertically aligned to a score bar. A curved arc below the bars links two dissimilar nodes.}
  \label{fig:alternative-tree-design}
\end{figure}

\crrev{\textbf{Design alternative.}
  We also implemented a top-down tree in which each node had a unique horizontal coordinate, with an aligned score bar below it and arcs between bars whose node dissimilarity exceeded a threshold (Fig.~\ref{fig:alternative-tree-design}).
  The threshold limited the visible arcs instead of showing links between every pair.
  During design iteration, E1 found the tree difficult to read when the population exceeded approximately 50 nodes because the tree, bars, and arcs competed for horizontal space.
We therefore selected the radial tree for the population overview.}

\subsection{Merge Recommendation}
\label{sec:appendix-merge-recommendation}

\crrev{Given a node $s$ selected for \feature{Merge}, \toolname first excludes incorrect, banned, and already selected nodes from the candidate partners.
  For each remaining candidate $c$, it computes a quality value $q(c)$ from the program's evaluation score and a dissimilarity value $d(s,c)=1-\cos(\mathbf{e}_s,\mathbf{e}_c)$ from the selected embedding source.
  Experts can use embeddings of either the source code or the generated reasoning, depending on whether they want to compare implementations or algorithmic strategies.
  The quality and dissimilarity values are separately min--max normalized to $\hat q(c)$ and $\hat d(s,c)$.
  Candidates from a different island receive a small dissimilarity preference of $0.1$, and candidates retained in the evolution archive receive a small quality preference of $0.05$; both normalized values remain capped at $1$.
  The archive is a bounded set of correct programs retained by the evolution engine for subsequent parent and inspiration selection.
The final recommendation score is}
\begin{equation}
  \crrev{r(c)=\lambda\hat d(s,c)+(1-\lambda)\hat q(c),}
  \label{eq:merge-recommendation}
\end{equation}
\crrev{where experts can adjust $\lambda$ to balance dissimilarity and quality; the default is $0.6$.
  \toolname presents the five highest-ranked candidates.
Candidates without the selected embedding are assigned a dissimilarity value of $0.5$ before normalization; when no usable embeddings are available, the method ranks candidates by quality alone.}

\crrev{This ranking does not determine whether two strategies are semantically complementary.
  It narrows the candidate set to programs that combine strong evaluation results with different implementations or algorithmic strategies.
The expert then inspects their scores, strategy similarity, and code differences before selecting a partner, and can provide optional natural-language guidance about which ideas the merged child should preserve or combine.}


\section{Task Details: OXC Routing Optimization}
\label{sec:appendix-oxc}

This appendix provides an overview of the task and the full mathematical formulation of the \gls{oxc} routing optimization task used in Section~\ref{sec:case-study}.

\subsection{Task Overview}

Adapted from the ICPC 2025 Online Winter Challenge\footnote{\url{https://icpc.global/regionals/finder/Winter-Challenge-2026}}, the task aims to optimize traffic routing in a large-scale AI computing cluster interconnected by \glspl{oxc}.
The goal is to evolve programs that jointly determine (1) the physical topology configuration of \glspl{oxc} and (2) traffic routing strategies over the configured topology.
Each program (evolution node) represents a candidate routing policy and is evaluated using a metric balancing two competing objectives: (1) minimizing peak link congestion and (2) minimizing the physical reconfiguration cost by \glspl{oxc} switching. Each program must complete execution within 50s during evaluation.
The search space is highly complex, as it involves combinatorial optimization over both network topology and traffic assignment.
Effective solutions require coordinating discrete topology decisions with continuous traffic distribution under performance constraints.
The task is challenging, as the program must jointly optimize topology configuration and traffic routing across a large-scale network, requiring strong algorithmic design skills to navigate the complex search space effectively.

\subsection{Mathematical Formulation}
\textbf{Network architecture.}
The multi-plane network comprises $N$ Groups, $M$ OXCs, $S$ Spines per Group, and $L$ Leaves per Group. To ensure scalability, the network is partitioned into $P$ independent planes, with $K$ parallel links between Spines and OXCs within the same plane. Cross-plane links are strictly prohibited. The total number of ports per OXC, $R$, is:
$$R = N \cdot (S/P) \cdot K$$

A specific port on an OXC is uniquely mapped by Group $i$, Spine $j$, and link $k$ as:
$$Idx_{port} = i \cdot (S/P) \cdot K + (j \bmod{S/P}) \cdot K + k$$

A valid routing path for real-time queries must symmetrically traverse the network: \textit{LeafA $\leftrightarrow$ SpineA $\leftrightarrow$ OXC $\leftrightarrow$ SpineB $\leftrightarrow$ LeafB}.

\textbf{Dual optimization objectives.}
Under strict temporal execution budgets, the system must dynamically balance two competing objectives:
\begin{enumerate}
  \item \textbf{Minimize Maximum Flow Conflict:} Alleviate network bottlenecks by minimizing the peak active flows $C(e)$ passing through any physical link $e \in E$:
    $$\min \left( \max_{e \in E} C(e) \right)$$
  \item \textbf{Minimize OXC Adjustment Cost:} Reduce the time penalty of physically rotating micro-mirrors, measured by the edit distance between successive port connection states $T_{t-1}$ and $T_t$:
    $$\min D(T_t, T_{t-1}) = |T_{t-1} \setminus T_t| + |T_t \setminus T_{t-1}|$$
\end{enumerate}


\section{Evaluation Study Details}
\label{sec:appendix-eval-details}

\subsection{Participants}
\label{sec:appendix-eval-participants}

\begin{table}[H]
  \caption{Evaluation study participants. LLM-Evo Exp. indicates self-reported experience with \gls{llm}-based evolutionary systems: N = None, B = Basic (understand concepts or short-term use), M = Medium (used similar systems with extended time), E = Expert (developing or researching). All participants are daily users of \gls{llm}-based programming tools.}
  \label{tab:eval-participants}
  \small
  \begin{tabular}{p{0.04\linewidth} p{0.22\linewidth} p{0.30\linewidth} p{0.14\linewidth}}
    \toprule
    \textbf{ID} & \textbf{Background} & \textbf{Domain} & \textbf{LLM-Evo Exp.} \\
    \midrule
    U1 & MSc Student & Physics & B \\
    U2 & PhD Student & \gls{dl} & N \\
    U3 & Software Engineer & Backend Optimization & B \\
    U4 & MSc Student & \gls{hci} & B \\
    U5 & MSc Student & Statistics & B \\
    U6 & PhD Student & \gls{dl} & B \\
    U7 & MSc Student & \gls{dl} & B \\
    U8 & MSc Student & \gls{dl} & M \\
    U9 & Software Engineer & Database Optimization & B \\
    U10 & MSc Student & \gls{dl} & B\\
    U11 & MSc Student & \gls{dl} & N \\
    U12 & MSc Student & Mechanical Engineering & B \\
    \bottomrule
  \end{tabular}
\end{table}

\subsection{Tasks}
\label{sec:appendix-eval-tasks}

\textbf{Task A: Drone Multi-Objective Path Optimization (MOP).}
A drone must fly from a fixed start point to an end point through a series of intermediate waypoints.
The evolvable function outputs waypoint coordinates that balance three competing objectives: flight path length, noise penalty from proximity to schools, and signal degradation from distance to base stations.

\textbf{Task B: Server Routing on Undirected Graphs (GraphMOP).}
In an edge computing network modeled as an undirected graph, tasks generated at terminal nodes must be routed to server nodes for processing.
The evolvable function determines server assignments and routing paths, balancing two competing objectives: end-to-end latency (transmission delay plus queuing delay) and system energy consumption.

Both tasks were chosen because they involve multi-objective optimization with non-trivial trade-offs, require domain reasoning to interpret nodes (not just numerical scores), and produce evolution trees with diverse algorithmic strategies amenable to expert analysis.
Each task\footnote{\href{https://shinka-mop.reify.ing}{Web link for ShinkaEvolve Task A} and \href{https://shinka-graph-mop.reify.ing}{Web link for ShinkaEvolve Task B}}$^{,}$\footnote{\href{https://evolvis-mop-03.reify.ing}{Web link for \toolname Task A} and \href{https://evolvis-graph-mop-03.reify.ing}{Web link for \toolname Task B}} was seeded with a pre-run evolution containing approximately 50 nodes across 6 islands.
In the \toolname condition, the evolution remained live, allowing participants to execute interventions and observe new nodes in real time.

\subsection{Counterbalancing}
\label{sec:appendix-eval-counterbalancing}

To control for ordering and task effects, we counterbalanced the assignment of conditions and tasks (\S~\ref{sec:appendix-eval-tasks}) using a $2 \times 2$ Latin square design:

\begin{table}[h]
  \caption{Counterbalanced assignment of conditions and tasks.}
  \label{tab:eval-counterbalancing}
  \small
  \centering
  \begin{tabular}{lll}
    \toprule
    \textbf{Group} & \textbf{Session 1} & \textbf{Session 2} \\
    \midrule
    G1 ($n=3$) & Baseline + Task A & \toolname + Task B \\
    G2 ($n=3$) & \toolname + Task A & Baseline + Task B \\
    G3 ($n=3$) & Baseline + Task B & \toolname + Task A \\
    G4 ($n=3$) & \toolname + Task B & Baseline + Task A \\
    \bottomrule
  \end{tabular}
\end{table}

\subsection{Measures}
\label{sec:appendix-eval-measures}

We collected four types of data after each condition:

\subsubsection{Custom Likert Questionnaire (7-point scale)}

The following items were shared across both conditions (1 = Strongly Disagree, 7 = Strongly Agree):

\textbf{Understandability (5 items, both conditions).}
\begin{enumerate}
  \item[\textbf{L1}] (\DR{3}) The system's visualization design allows me to clearly understand how the entire code population evolves over generations.
  \item[\textbf{L2}] (\DR{1}) The system allows me to easily discern the overall algorithmic strategies adopted by the LLM across different branches.
  \item[\textbf{L3}] (\DR{4}) The system helps me easily locate code that exhibits abnormal or special performance.
  \item[\textbf{L4}] (\DR{2}) The system enables me to effectively compare different solutions and understand the differences between them.
  \item[\textbf{L5}] (\DR{1}) The system allows me to clearly see the algorithm used by each node.
\end{enumerate}

\textbf{Usability (3 items, both conditions).}
\begin{enumerate}
  \item[\textbf{L6}] The system is easy to use for me.
  \item[\textbf{L7}] The information provided by the system is organized very clearly on the screen, and I can easily find what I need.
  \item[\textbf{L8}] The system information effectively helps me complete my tasks and operational scenarios.
\end{enumerate}

The following items were collected only in the \toolname condition:

\textbf{Maestro Chat (1 item, \toolname only).}
\begin{enumerate}
  \item[\textbf{L9}] AI-assisted code explanations (like Maestro Chat) helped me understand solutions that would be difficult for me to grasp on my own.
\end{enumerate}

\textbf{Intervenability (3 items, \toolname only).}
\begin{enumerate}
  \item[\textbf{L10}] (\DR{5}) The system enables me to accurately identify where intervention is needed in the current evolutionary tree.
  \item[\textbf{L11}] (\DR{6}) The system allows me to flexibly control the evolution pace, pausing to observe or stepping through as needed.
  \item[\textbf{L12}] (\DR{5}) The system effectively supports me in conducting effective interventions within the evolutionary tree.
\end{enumerate}

\subsubsection{NASA-TLX}
The NASA Task Load Index~\cite{hart_development_1988} was administered after each condition (both conditions, 7-point scale where 1 = Very Low and 7 = Very High; for Performance, 1 = Good and 7 = Poor):
\begin{enumerate}
  \item Mental Demand: How mentally demanding was the task?
  \item Physical Demand: How physically demanding was the task?
  \item Temporal Demand: How hurried or rushed was the pace of the task?
  \item Performance: How successful were you in accomplishing what you were asked to do?
  \item Effort: How hard did you have to work to accomplish your level of performance?
  \item Frustration: How insecure, discouraged, irritated, stressed, and annoyed were you?
\end{enumerate}

\subsubsection{Open-ended Questions}

The following questions were asked in both conditions:
\begin{enumerate}
  \item[\textbf{O1}] Please describe in detail how you identified the well-performing branch from the evolutionary tree or scatter plot. Which visual cues provided the most critical guidance?
  \item[\textbf{O2}] How did you identify the first critical code variant that triggered the issue in this branch?
  \item[\textbf{O3}] How did you understand the algorithm logic of code variants through the interface?
  \item[\textbf{O4}] What was the most cognitively demanding part of the experiment?
  \item[\textbf{O5}] At what moment did the system make you feel like you had a clear global view?
  \item[\textbf{O6}] In this human-AI co-evolution process, do you feel your sense of control over the algorithmic direction was strengthened or weakened?
\end{enumerate}

The following questions were asked only in the \toolname condition:
\begin{enumerate}
  \item[\textbf{O7}] Please explain the design rationale behind your intervention prompt. Based on which specific bottleneck or logical flaw did you decide to issue this instruction to the LLM?
  \item[\textbf{O8}] Did you use Maestro Chat to understand code or code differences? If so, describe a specific moment when it helped. If not, why not?
\end{enumerate}

\subsubsection{Post-study Interview}
A semi-structured interview after both conditions captured preference comparisons, feature-level feedback, and reflections on how each tool shaped understanding.

\subsection{Procedure}
\label{sec:appendix-eval-procedure}

Each evaluation session lasted approximately 80 minutes and followed this protocol:

\begin{enumerate}
  \item \textbf{Introduction and consent (5 min).}
    We explained the study goals and obtained informed consent.

  \item \textbf{Tutorial on \gls{llm}-driven evolution (5 min).}
    Using the same visual handout from the formative study (Appendix~\ref{sec:appendix-fs-procedure}), we introduced the evolutionary loop, reasoned mutations, and the island model.

  \item \textbf{Condition 1 --- Tool training (5 min).}
    We demonstrated the assigned interface's key features.

  \item \textbf{Condition 1 --- Tasks (25 min).}
    Participants explored the evolution result following a structured task sheet (in the \toolname condition, the evolution was live, allowing real-time interventions):
    \begin{itemize}
      \item \emph{Sensemaking tasks (15 min):}
        Participants were asked to identify the three most promising nodes and explain why, describe the major algorithmic strategies that emerged during evolution, identify any nodes exploiting evaluation function weaknesses, and find a low-scoring node with a promising idea worth developing further.
      \item \emph{Intervention tasks (10 min):}
        In the \toolname condition, participants selected a node and wrote a \gls{nl} suggestion to guide the next mutation, then selected 2--3 nodes and initiated a multi-parent merge.
        In the baseline condition, participants instead wrote down what modifications they would make and which nodes they would combine, capturing intervention \emph{intent} without tool support for qualitative comparison.
    \end{itemize}

  \item \textbf{Condition 1 --- Questionnaire (5 min).}
    Participants completed the custom Likert questionnaire (Understandability, Usability, and, for \toolname, Intervenability and Maestro Chat items) and open-ended questions, followed by the NASA-TLX.

  \item \textbf{Break (5 min).}

  \item \textbf{Condition 2 (35 min).}
    Steps 3--5 repeated with the other interface and task.

  \item \textbf{Post-study interview (10 min).}
    A semi-structured interview covering preference comparison between the two interfaces, feature-level feedback on \toolname, suggestions for missing features, and how participants' understanding of the evolution differed between conditions.
\end{enumerate}

\subsection{Detailed Results}
\label{sec:appendix-eval-results}

\begin{table}[H]
  \caption{Paired comparison of shared Likert items (7-point scale; higher is better). All differences are significant at $p < 0.05$ (Wilcoxon signed-rank test). Effect sizes: $|\delta| < 0.33$ = small, $< 0.474$ = medium, $\geq 0.474$ = large.}
  \label{tab:likert-results}
  \small
  \begin{tabular}{p{0.32\linewidth} r r r r}
    \toprule
    \textbf{Item} & \textbf{Ours} & \textbf{Base} & \textbf{$p$} & \textbf{$\delta$} \\
    & \textbf{Med} & \textbf{Med} & & \\
    \midrule
    \multicolumn{5}{l}{\emph{Understandability}} \\
    Population evolution (\DR{3}) & 6.0 & 5.0 & .016 & .451\textsuperscript{m} \\
    Branch strategies (\DR{1}) & 6.0 & 5.0 & .039 & .403\textsuperscript{m} \\
    Locate abnormal (\DR{4}) & 7.0 & 5.0 & .023 & .521\textsuperscript{L} \\
    Compare solutions/nodes (\DR{2}) & 6.0 & 5.0 & .020 & .486\textsuperscript{L} \\
    Node algorithm (\DR{1}) & 6.0 & 5.0 & .031 & .368\textsuperscript{m} \\
    \midrule
    \multicolumn{5}{l}{\emph{Usability}} \\
    Easy to use & 6.0 & 4.5 & .027 & .472\textsuperscript{m} \\
    Info organized & 7.0 & 5.0 & .016 & .639\textsuperscript{L} \\
    Task effective & 7.0 & 5.0 & .002 & .694\textsuperscript{L} \\
    \bottomrule
  \end{tabular}

  \raggedright\scriptsize\textsuperscript{m}Medium effect. \textsuperscript{L}Large effect.
\end{table}

\begin{table}[H]
  \caption{NASA-TLX results (7-point scale; lower is better). Significant differences are marked with *. \crrev{The Performance subscale measures participants' perceived success in completing the study task, not generated-program performance.}}
  \label{tab:tlx-results}
  \small
  \begin{tabular}{p{0.25\linewidth} r r r r}
    \toprule
    \textbf{Subscale} & \textbf{Ours} & \textbf{Base} & \textbf{$p$} & \textbf{$\delta$} \\
    & \textbf{Med} & \textbf{Med} & & \\
    \midrule
    Mental Demand & 3.5 & 5.5 & .043* & $-$.347\textsuperscript{m} \\
    Physical Demand & 2.0 & 3.0 & .070 & $-$.354\textsuperscript{m} \\
    Temporal Demand & 3.0 & 4.5 & .016* & $-$.479\textsuperscript{L} \\
    Performance & 3.0 & 3.0 & .992 & .056 \\
    Effort & 3.5 & 3.5 & .254 & $-$.174 \\
    Frustration & 2.0 & 4.0 & .004* & $-$.535\textsuperscript{L} \\
    \midrule
    \textbf{Overall} & \textbf{3.08} & \textbf{4.25} & \textbf{.004*} & $-$\textbf{.417}\textsuperscript{m} \\
    \bottomrule
  \end{tabular}

  \raggedright\scriptsize\textsuperscript{m}Medium effect. \textsuperscript{L}Large effect.
\end{table}

\end{document}